\documentclass[preprint,12pt]{elsarticle}

\usepackage{amssymb}
\usepackage{amsthm}
\newdefinition{sRemark}{Software Remark}

\journal{}

\usepackage{amsmath}
\allowdisplaybreaks 

\usepackage{url} 
\usepackage{rotating} 

\usepackage{pgf,tikz} 
\usetikzlibrary{arrows, arrows.meta, bending}
\usepackage[final]{pdfpages} 

\newcommand{\VS}{\operatorname{VS}}
\newcommand{\difft}[1]{\dot{#1}}

\def \R{\mathbb{R}}
\def \Z{\mathbb{Z}}

\newcommand{\msub}{\mathord{/\kern-3pt/}}

\newlength{\lrtskip}
\begin{document}

\begin{frontmatter}





\title{Identifying the Parametric Occurrence of Multiple Steady States for some Biological Networks}

\author[BATH]{R.~Bradford}
\ead{R.J.Bradford@bath.ac.uk}
\author[BATH]{J.H.~Davenport}
\ead{J.H.Davenport@bath.ac.uk}
\author[COV]{M.~England\corref{cor1}}
\ead{Matthew.England@coventry.ac.uk}
\author[BONN]{H.~Errami}
\ead{errami@cs.uni-bonn.de}
\author[JINR]{V.~Gerdt}
\ead{gerdt@jinr.ru}
\author[CNRS]{D.~Grigoriev}
\ead{dmitry.grigoryev@univ-lille.fr}
\author[BIT]{C.~Hoyt}
\ead{cthoyt@gmail.com}
\author[SAoS]{M.~Ko{\v s}ta}
\ead{marek.kosta@savba.sk}
\author[MONT]{O.~Radulescu}
\ead{ovidiu.radulescu@umontpellier.fr}
\author[STURMF,STURMD]{T.~Sturm}
\ead{thomas@thomas-sturm.de}
\author[BONN]{A.~Weber}
\ead{weber@cs.uni-bonn.de}


\address[BATH]{Department of Computer Science, University of Bath, UK}
\address[COV]{Faculty of Engineering, Environment and Computing, Coventry University, UK}
\address[BONN]{Institute for Informatics, University of Bonn, Germany}
\address[JINR]{Joint Institute for Nuclear Research (JINR), Dubna, Russian Federation\\ and Friendship University of Russia (RUDN University), Moscow, Russian Federation}
\address[CNRS]{CNRS \& University of Lille, France}
\address[BIT]{Department of Life Science Informatics, B-IT, University of Bonn, Germany}
\address[SAoS]{Slovak Academy of Sciences, Slovakia}
\address[MONT]{DIMNP, University of Montpellier, France}
\address[STURMF]{CNRS, Inria, and the University of Lorraine, Nancy, France}
\address[STURMD]{MPI Informatics and Saarland University, Saarbr\"{u}cken, Germany}

\cortext[cor1]{Corresponding author}

\begin{abstract}
We consider a problem from biological network analysis of determining regions in a parameter space over which there are multiple steady states for positive real values of variables and parameters.  We describe multiple approaches to address the problem using tools from Symbolic Computation.  We describe how progress was made to achieve semi-algebraic descriptions of the multistationarity regions of parameter space, and compare symbolic results to numerical methods.

\newpage

The biological networks studied are models of the mitogen-activated protein kinases (MAPK) network which has already consumed considerable effort using special insights into its structure of corresponding models.  Our main example is a model with 11 equations in 11 variables and 19 parameters, 3 of which are of interest for symbolic treatment.  The model also imposes positivity conditions on all variables and parameters.  

We apply combinations of symbolic computation methods designed for mixed equality/inequality systems, specifically virtual substitution, lazy real triangularization and cylindrical algebraic decomposition, as well as a simplification technique adapted from Gaussian elimination and graph theory.

We are able to determine multistationarity of our main example over a 2-dimensional parameter space.  We also study a second MAPK model and a symbolic grid sampling technique which can locate such regions in 3-dimensional parameter space.
\end{abstract}


\end{frontmatter}


\section{Introduction}
\label{SEC:Intro}

In this work we describe the application of combinations of symbolic computation methods in various computer algebra systems to a key problem from computational biology.  The work serves to demonstrate how recent advances in such algorithms, and crucially their effective combination, allows for their application on problem instances previously thought beyond reach.  In this introduction we start by describing the biological networks that are our topic of study, and highlight previous relevant work.  We then outline the remainder of the paper and clarify the relationship of this article to prior work.

\subsection{Multistationarity}
\label{SUBSEC:BioIntro}

The mathematical modelling of intra-cellular biological processes has been using nonlinear ordinary differential equations since the early ages of mathematical biophysics in the 1940s and 50s \citep{rashevsky1960mathematical}.   A standard modelling choice for cellular circuitry is to use chemical reactions with mass action law kinetics, leading to polynomial differential equations.   Rational  functions kinetics, for instance the Michaelis-Menten kinetics, can generally be decomposed into several mass action steps.

An important property of biological systems is their \emph{multistationarity} by which we mean their having multiple stable steady states.  It is instrumental to cellular memory and cell differentiation during development or regeneration of multicellular organisms and is also used by micro-organisms in survival strategies.   

It is thus important to determine the parameter values for which a biochemical model is multistationary. As demonstrated in the next section, with mass action reactions, testing for multiple steady states boils down to counting real positive solutions of algebraic systems and so is suitable for study with Symbolic Computation and Computer Algebra Systems.

The models studied in this paper concern intracellular signaling pathways. These pathways transmit information about the cell environment by inducing cascades of protein modifications (phosphorylation) all the way from the plasma membrane via the cytosol to genes in the cell nucleus. Multistationarity of signaling usually occurs as a result of activation of upstream signaling proteins by downstream components \citep{BhallaLyengar99}. A different mechanism for producing multistationarity in signaling pathways was proposed by \citet{Markevich2004}. In this mechanism the cause of multistationarity are multiple phosphorylation/ dephosphorylation cycles that share enzymes. A simple, two steps phosphorylation/dephosphorylation cycle is capable of ultrasensitivity, a form of all or nothing response with no multiple steady states (the Goldbeter--Koshland mechanism). In multiple
phosphorylation/dephosphorylation cycles, enzyme sharing provides competitive interactions and positive feedback that ultimately leads to multistationarity \citep{Markevich2004,legewie2007competing}.

\subsection{Bistability}
\label{SUBSEC:BiStabDetails}

Multistationarity has important consequences on the capacity of signaling pathways to process biological signals, even in its elementary form of two stable steady states.  This is known as \emph{bistability} and is present in our case study problems.  Bistable switches can act as memory circuits storing the information needed for later stages of processing \citep{WengBhallaLyengar1999}.  The response of bistable signaling pathways shows hysteresis, namely dynamic and static lags between input and output. Because of hysteresis one can have, at the same time, a sharp binary response and protection against chatter noise.

\subsection{Prior Symbolic Work}

Our study is complementary to works applying numerical methods to ordinary
differential equations models used for biology applications. \citet{gross2016numerical} used polynomial homotopy continuation methods for
global parameter estimation of mass action models. Bifurcations and
multistationarity of signaling cascades was studied with numerical methods based
on the Jacobian matrix by \citet{zumsande2010bifurcations}. 

Algorithmically the task will be to count the positive real solutions of a parameterised
system of polynomial or rational systems, making symbolic methods a possible tool.
Due to the high computational complexity of this task \citep{GrigorevVorobjov1988a}  considerable work has been done to use specific properties of networks and to investigate the potential of multistationarity of a biological network  out of the network structure.  

This only determines whether or not there exist rate constants allowing multiple steady states, instead of coming up 
with a semi-algebraic description of the range of parameters yielding this property. These approaches can be traced back to the origins of 
Feinberg's {\em Chemical Reaction Network Theory} (CRNT) whose main result is 
that networks of deficiency 0 have a unique positive steady state for all rate constants \citep{Feinberg1987,CraciunDickensteinShiuSturmfels2009}. 
We refer to \citet{Conradi2008, PerezMillan2015, Johnston2014}, and \citet{Conradi2017} for the use of CRNT and other graph theoretic methods to determine potential existence of multiple positive steady states, with \citet{JoshiShiu2015} giving a survey.  

Given a bistable mechanism it is also important to compute the bistability domains in parameter space: the parameter values for which there is more than one stable steady state.
The size of bistability domains gives the spread of the hysteresis and quantifies the robustness of the switches. The work of \citet{WangXia2005a} is relevant here: they used symbolic tools, including cylindrical algebraic decomposition as we do, to determine the number of steady states and their stability for several systems.  They reported results up to a 5-dimensional system using specified parameter values, but their method is extensible to parametric questions. 
Higher-dimensional systems were studied using sign conditions on the coefficients of the characteristic polynomial of the Jacobian. In some cases these guarantee uniqueness of the steady state \citep{ConradiMincheva}.

\newpage

\subsection{Outline and New Contributions}
\label{SUBSEC:Outline}

In Section \ref{SEC:Problem} we outline the particular biological model and symbolic problem that we aim to solve: BioModel 26 of the MAPK network, which can be found as Model 26 in the BioModels Database of \citep{Li2010a}.  

In Sections \ref{SEC:Redlog} and \ref{SEC:Maple} we describe two independent symbolic attempts to solve the problem.  The first in Section \ref{SEC:Redlog} is able to identify symbolically the multistationarity regions of a 1-dimensional parameter space with a combination of Virtual Substitution and Cylindrical Algebraic Decomposition in the Redlog package for Reduce.  The second in Section \ref{SEC:Maple} goes on to give full semi-algebraic solution formulae with a combination of Real Triangularization and Cylindrical Algebraic Decomposition using the Regular Chains Library for Maple.  The solutions were obtained in different computer algebra systems using different fundamental algorithms, but all from the family of methods for real quantifier elimination.
We move on in Section \ref{SEC:Preproc} to describe a new pre-processing method for the problems inspired by graph theory and Gaussian elimination.  Then in Section \ref{SEC:New2Para} we describe how a combination of ideas from all three preceding sections can be combined to provide solutions over a 2-dimensional parameter space.

In Section \ref{SEC:Stability} we discuss testing the stability of fixed points.  Then in Section \ref{SEC:Mod28} we consider an alternative larger model from the MAPK network (Model 28 in the BioModels Database of \citep{Li2010a}).  In Section \ref{SEC:vsNumerical} we compare the models and detour to describe a symbolic grid sampling approach to this problem, including a comparison of this to a leading numerical solver.  
We consider how further progress could be achieved in Section \ref{SEC:Further}, identifying a conjecture for determining where multistationarity for MAPK may occur without the costly calculations described.  Finally we summarise and give final thoughts in Section \ref{SEC:Final}.

This journal article follows published conference works at ISSAC 2017 \citep{BioISSAC17} and CASC 2017 \citep{BioCASC17}.  The present article reproduces this material clarifying, correcting and extending in places.  In particular, Sections \ref{SEC:Redlog} and \ref{SEC:Maple} were largely described in the ISSAC 2017 paper and Sections \ref{SEC:Preproc} and \ref{SEC:vsNumerical} in the CASC 2017 paper.  The most notable new contributions are given in Section \ref{SEC:New2Para}, where we describe for the first time semi-algebraic solutions with two free parameters; and in Section \ref{SEC:Further}, where we identify a promising conjecture for future investigation.  

\newpage

\section{Problem Outline}
\label{SEC:Problem}

\subsection{MAPK Bio-Model 26}
\label{SUBSEC:Mod26}

The model of the MAPK cascade we are investigating can be found in the BioModels Database \citep{Li2010a} as Model 26\footnote{\url{www.ebi.ac.uk/biomodels-main/BIOMD0000000026}}.  This is the first version of the models proposed by \citet{Markevich2004} corresponding to the so-called distributive ordered phosphorylation/dephosphorylation mechanism.  Hereafter we will refer to it as Model 26.

It is given by the following set of differential equations.
We have renamed the species names to $x_1$, \dots,~$x_{11}$ and the rate
constants to $k_1$, \dots,~$k_{16}$ to facilitate reading.  As usual $\difft{x}$ means the time derivative of $x$.
\begin{align}
  \difft{x_1} ={} &  k_{2} x_{6} + k_{15} x_{11} - k_{1} x_{1} x_{4} - k_{16} x_{1} x_{5}\nonumber\\
  \difft{x_2} ={} &  k_{3} x_{6} + k_{5} x_{7} + k_{10} x_{9} + k_{13} x_{10} - 
        x_{2} x_{5} (k_{11} + k_{12}) - k_{4} x_{2} x_{4}\nonumber\\
  \difft{x_3} ={} &  k_{6} x_{7} + k_{8} x_{8} - k_{7} x_{3} x_{5}\nonumber\\
  \difft{x_4} ={} &  x_{6} (k_{2} + k_{3}) + x_{7} (k_{5} + k_{6}) - k_{1} x_{1} x_{4} - k_{4} x_{2} x_{4}\nonumber\\
  \difft{x_5} ={} &  k_{8} x_{8} + k_{10} x_{9} + k_{13} x_{10} + k_{15} x_{11} - \nonumber\\
    &   \quad  x_{2} x_{5} (k_{11} + k_{12}) - k_{7} x_{3} x_{5} - k_{16} x_{1} x_{5}\nonumber\\
  \difft{x_6} ={} &  k_{1} x_{1} x_{4} - x_{6} (k_{2} + k_{3})\nonumber\\
  \difft{x_7} ={} &  k_{4} x_{2} x_{4} - x_{7} (k_{5} + k_{6})\nonumber\\
  \difft{x_8} ={} &  k_{7} x_{3} x_{5} - x_{8} (k_{8} + k_{9})\nonumber\\
  \difft{x_{9}} ={} &  k_{9} x_{8} - k_{10} x_{9} + k_{11} x_{2} x_{5}\nonumber\\
  \difft{x_{10}} ={} &    k_{12} x_{2} x_{5} - x_{10} (k_{13} + k_{14})\nonumber\\
  \difft{x_{11}} ={} &    k_{14} x_{10} - k_{15} x_{11} + k_{16} x_{1} x_{5}.\label{EQ:thesystem}
\end{align}
Later, we will use $\overline{(\ref{EQ:thesystem})}$ to refer to (\ref{EQ:thesystem}) with all the left hand sides replaced by $0$ in order to find fixed points of the system.
The BioModels Database gives us meaningful values for the
rate constants:
\begin{align}
  k_{1} &= 0.02,&\!
  k_{2} &= 1,&\!
  k_{3} &= 0.01,&\!
  k_{4} &= 0.032,\nonumber\\
  k_{5} &= 1,&\!
  k_{6} &= 15,&\!
  k_{7} &= 0.045,&\!
  k_{8} &= 1,\nonumber\\
  k_{9} &= 0.092,&\!
  k_{10} &= 1,&\!
  k_{11} &= 0.01,&\!
  k_{12} &= 0.01,\nonumber\\
  k_{13} &= 1,&\!
  k_{14} &= 0.5,&\!
  k_{15} &= 0.086,&\!
  k_{16} &= 0.0011.\label{EQ:rcestimates}
\end{align}
Some of these values are accurately measured and some are \emph{well-educated guesses}.  For the purpose of our study we assume they are all suitable.

We may add three linear conservation constraints to this system, which in turn introduce three further constant parameters $k_{17}$, $k_{18}$, $k_{19}$:
\begin{align}
  x_{5}  + x_{8} + x_{9} + x_{10} + x_{11} &= k_{17}\nonumber\\
  x_{4} + x_{6} + x_{7} &=  k_{18}\nonumber\\
  x_{1} + x_{2} + x_{3} + x_{6} + x_{7} + x_{8} + x_{9} + x_{10} + x_{11} &=  k_{19}.\label{EQ:claws}
\end{align}
Computations to produce these, for example in MathWorks SimBiology, use the left-null space of the stoichiometric matrix under positivity conditions.  For details see for example \citet{schuster1991determining}.

The constants $k_{17}, k_{18}$, and $k_{19}$ represent total initial concentrations of cell substances, and meaningful values are harder to obtain than for (\ref{EQ:rcestimates}).  The following are some realistic value estimates, used by \citet{Markevich2004}:
\begin{align}
  k_{17} &= 100,&  k_{18} &= 50,& k_{19} \in [200,500].\label{EQ:clestimates}
\end{align}
These 
should be considered significantly less reliable than those in (\ref{EQ:rcestimates}).  Indeed, the long-term goal of our research is to treat all three of these together parametrically, although in the present work we produce results only with $0-2$ of these parameters free.

Our computational biology problem is to identify regions in $(k_{17}, k_{18}, k_{19})$ parameter space over which the system formed by the unions of constraints in (\ref{EQ:thesystem}) and (\ref{EQ:claws}) under estimates (\ref{EQ:rcestimates}) exhibits multistationarity.

The system has several special structure properties, e.g.  it is a so called MESSI system \citep{Millan2016a}. However, in the following we will not directly use this structure property.
The non-linearities occurring in the system are at most quadratic. As by introducing new variables the general polynomial case can be reduced to such a case and from a dynamical systems perspective point of view
already quadratic systems are capable to generate all kinds of structurally stable dynamics including chaos \citep{Vakulenko2015} this property is not restrictive.

\subsection{Real Algebraic Problem}
\label{SUBSEC:SymProb}

To identify fixed points we formulate a real algebraic problem by first replacing the left hand sides of all equations in (\ref{EQ:thesystem}) with $0$, which as noted above we denote $\overline{(\ref{EQ:thesystem})}$.  This, together with the equations in (\ref{EQ:claws}), yields an algebraic system with polynomials in
\begin{displaymath}
F\subset\Z[k_1,\dots,k_{19}][x_1,\dots,x_{11}].
\end{displaymath}
However, ideal theory is not sufficient, as we are concerned only with real valued solutions.  Further, we have the additional inequality restrictions that all entities in our model are strictly positive.  This yields an additional system
\[
P=\{k_{1},\dots,k_{19},x_1,\dots,x_{11}\}\subset\Z[k_1,\dots,k_{19}][x_1,\dots,x_{11}]
\]
establishing a side condition on the solutions of $F$ that all variables $x_i$ and parameters $k_i$ of $P$ be positive.  In terms of first-order logic our specification of $F$ and $P$
yields a quantifier-free Tarski formula,
\begin{equation}
\label{eq:varphi}
\varphi=\bigwedge_{f\in F}f=0\land\bigwedge_{v\in P}v>0.
\end{equation}
The estimations for the rate constants in (\ref{EQ:rcestimates}) formally
establish a substitution rule $\sigma = [0.02/k_1, \dots, 0.0011/k_{16}]$ in postfix notation, which can be applied to $F$, $P$, or $\varphi$.  Applying this to $\varphi$; converting the floats from (\ref{EQ:rcestimates}) into rational numbers; and multiplying over common denominators, gives us the quantifier-free Tarski formula $\psi$ below.  
\begin{align}
\psi &= -200x_{1}x_{4}-11x_{1}x_{5} + 860x_{11} + 10000x_{6} = 0
\nonumber \\
&\quad \land -16x_{2}x_{4}-10x_{2}x_{5} + 500x_{10} + 5x_{6} + 500x_{7} + 500x_{9} = 0 
\nonumber \\
&\quad \land -9x_{3}x_{5} + 3000x_{7} + 200x_{8} = 0 
\nonumber \\
&\quad \land -10x_{1}x_{4}-16x_{2}x_{4} + 505x_{6} + 8000x_{7} = 0 
\nonumber \\
&\quad \land -11x_{1}x_{5}-200x_{2}x_{5}-450x_{3}x_{5} + 10000(x_{8} + x_{9} + x_{10}) + 860x_{11} = 0 
\nonumber \\
&\quad \land 2x_{1}x_{4}-101x_{6} = 0 
\nonumber \\
&\quad \land 4x_{2}x_{4}-2000x_{7} = 0 
\nonumber \\
&\quad \land 45x_{3}x_{5}-1092x_{8} = 0 
\nonumber \\
&\quad \land 5x_{2}x_{5} + 46x_{8}-500x_{9} = 0 
\nonumber \\
&\quad \land x_{2}x_{5}-150x_{10} = 0 
\nonumber \\
&\quad \land 11x_{1}x_{5} + 5000x_{10}-860x_{11} = 0 
\nonumber \\
&\quad \land -k_{17} + x_{10} + x_{11} + x_{5} + x_{8} + x_{9} = 0 
\nonumber \\
&\quad \land -k_{18} + x_{4} + x_{6} + x_{7} = 0 
\nonumber \\
&\quad \land -k_{19} + x_{1} + x_{10} + x_{11} + x_{2} + x_{3} + x_{6} + x_{7} + x_{8} + x_{9} = 0 
\nonumber \\
&\quad \land x_{1}>0 \land x_{2}>0 \land x_{3}>0 \land x_{4}>0 \land x_{5}>0 
\nonumber \\
&\quad \land x_{6}>0 \land x_{7}>0  \land x_{8}>0  \land x_{9}>0 \land x_{10}>0 \land x_{11}>0 
\nonumber \\
&\quad \land k_{17}>0 \land k_{18}>0 \land k_{19}>0. \label{eq:3ParaTarski}
\end{align}

Our problem in real algebra is to obtain a semi-algebraic description of the regions in $(k_{17}, k_{18}, k_{19})$ parameter-space where there are multiple solutions of (\ref{eq:3ParaTarski}).  The multistationarity problem would also require to know about the stability of these solutions, as discussed in Section \ref{SEC:Stability}.

\subsection{Suitable Symbolic Technology}

This real algebraic problem is amenable to technology developed for real quantifier elimination.  Note that the number of indeterminates (variables and parameters) is high compared to those usually tackled by such technology.  However, the degrees involved are low, with every monomial at most degree 2, which helps make it tractable.  

As we will not include a priori information about the stability of the fixed points, we must not only consider the existence of (at least) two stable fixed points but also unstable fixed points.  Hence we simply investigate where in parameter space there exist multiple different roots $\mathbf{x}\in (0, \infty)^{11}$ of $F$.

In theory, any \emph{Real Quantifier Elimination} (QE) technology can directly handle the parametric existence of steady states, taking as input $\exists x_1\dots\exists x_{11}\varphi$ and producing as output a quantifier free formula in the parameters describing where solutions exists.  However, this is not sufficient to solve our problem as we are not only interested in the existence but also in the number of solutions.  We can use a specific QE tool to do this: Cylindrical Algebraic Decomposition.

\subsubsection{Cylindrical algebraic decomposition and its terminology}
\label{SSSEC:CAD}

\emph{Cylindrical Algebraic Decomposition} (CAD) was first proposed by Collins in the 1970s.  This original algorithm\footnote{see for example the work of \citet{ACM:84}.}  took as input a set of polynomials in $\mathbb{Z}[x_1, \dots, x_N]$, producing as output a set of \emph{cells} which together give a decomposition of $\mathbb{R}^n$ which is \emph{sign-invariant}, meaning each input polynomial has constant sign over each cell.  The sign-invariance means that the polynomials may be studied over an an infinite domain by querying a finite number of sample points: one per cell.

The cells are all semi-algebraic, meaning they can be described by a polynomial system, and they are arranged \emph{cylindrically}, meaning their projections with respect to a stated variable ordering are either equal or disjoint.  
The cylindricity means the semi-algebraic descriptions are triangular and the cells form cylinders over another (induced) CAD of $\mathbb{R}^{n-1}$ given by the projection of the $n$-dimensional cells.  All cells are either \emph{sections}, defined by a polynomial vanishing; or a \emph{sector}, defined as the space between two sections, or possibly extending infinitely.

Collins' algorithm proceeded with a system of: projection, which identified key polynomials in fewer variables; and lifting, where the induced CADs are incrementally constructed via substitution of sample points and univariate root isolation.  The act of projection must be defined so that working at a sample point may be concluded representative for the entire cell.  

There has been numerous extensions and improvements to CAD since Collins' original method.  The collection edited by \citet{CJ98} is a key resource; in particular the survey paper within by \citet{Collins1998}.  A more recent survey was given in the Introduction section of the work by \citet{BDEMW16}.  
A key choice for CAD is the variable ordering which defines the cylindricity property and controls the order steps are taken by the algorithm.  For use in  quantifier elimination CAD must project variables in the order they are quantified.  Our problem (\ref{eq:3ParaTarski}) is not quantified but our desire to understand the problem over parameter space means that we must project variables before parameters.  However, besides this the choice is free for us.  We define the \emph{main variable} of a polynomial / constraint to be the highest one present (first to be projected) in the ordering.

The worst-case time complexity of CAD is doubly exponential.  Traditionally, this is doubly exponential in the number of indeterminates, which would include our symbolically treated parameters.  However recent progress on CAD in the presence of  equational constraints (see for example the work of \citet{EBD15}), of which there are many in (\ref{eq:3ParaTarski}), allows us to conclude it is actually doubly-exponential in the number of variables minus the number of equational constraints at different levels of the projection \citep{ED16a}.  
Despite this, the number of variables present in (\ref{eq:3ParaTarski}) is too large for contemporary CAD implementations to tackle alone.  

\subsubsection{Combing with other symbolic tools}

We are able to make progress by combining CAD with additional symbolic methods.  Two independent investigations were undertaken.  The first, described in Section \ref{SEC:Redlog}, uses the Redlog package in Reduce and combines CAD with virtual substitution.  The second, described in Section \ref{SEC:Maple}, uses the Regular Chains Library in Maple and combines CAD with real triangularization.  In both cases we have combined the corresponding methods by hand, but automation is clearly possible.

\section{Using Real Quantifier Elimination Technology in Redlog}
\label{SEC:Redlog}

In this section we are going to combine \emph{Virtual Substitution} (VS) with CAD. The former smoothly eliminates the majority of the quantifiers while the latter allows us to count numbers of solutions via decomposition of the remaining low-dimensional spaces. 
That combination of methods requires the solution of several QE runs with each problem and some combinatorial arguments. Throughout this section we are performing computations using the Redlog Package~\citep{DolzmannSturm:97a} for Reduce revision r3606. Timings are reported for a 2.4 GHz Intel Core i7 with 3 GB RAM or cores on a compute server with similar speed and memory limitations.

\subsection{Virtual Substitution}
\label{SE:vsintro}

Substitution methods for quantifier elimination date back to an article from \citet{Weispfenning:88}, which treated the special case with only linear occurrences of the quantified variables. Originally motivated by the proof of tight complexity bounds for the real decision problem, that approach turned out to be applicable to practical problems, especially with many parameters. Consequently, the method was systematically generalized by Weispfenning and his students to arbitrary but bounded degrees \citep{Weispfenning:97,Weispfenning94a,Kosta:16a}.

Quantifier elimination proceeds from the inside to the outside of a prenex quantifier block. An innermost existential quantifier is eliminated by equivalently replacing it with a finite disjunction:
\[
\VS(\exists x_n\varphi):=\bigvee_{t\in E}\varphi[t\msub x_n],
\]
where $E$ is a finite \emph{elimination set} containing abstract \emph{test
  points $t=(\gamma,z)$}. The terms $z$ are derived from symbolic
representations of formal zeros of parametric univariate polynomials from
$\Z[x_1,\dots,x_{n-1}][x_n]$ occurring in $\varphi$ with possibly adding
infinitesimals $\pm\varepsilon$. They are guarded by quantifier-free formulas
$\gamma(x_1,\dots,x_{n-1})$ that guarantee the existence of the zeros in terms
of the parameters. Recall that regular term substitution maps terms to terms,
which naturally generalizes to corresponding maps on quantifier-free formulas.
\emph{Virtual substitution} $[t\msub x_n]$, in contrast, maps atomic formulas to
quantifier-free formulas. This allows to express the substitution of the terms
$z$ without using any non-standard symbols. Furthermore, virtual substitution
adds the guarding conditions $\gamma$ in a suitable way. For examples and
surveys of the virtual substitution method see the work of \citet{Sturm:17a,
  Sturm:18a}.

\subsection{Parameter Free Computations} 
\label{SUBSEC:RedlogParaFree}

We start by considering the case where all parameters in (\ref{eq:varphi}) are substituted for their estimates in (\ref{EQ:rcestimates}) and (\ref{EQ:clestimates}) (interpreted as rational numbers):
\[
\varphi_{500}=\varphi\sigma[100/k_{17},50/k_{18},500/k_{19}].
\]
The closed formula
$\bar\varphi_{500}=\exists x_1\dots\exists x_{11}\varphi_{500}$ states the
existence of a suitable real solution. In a first step, we solve for
$i\in\{1,\dots,11\}$ the following eleven QE problems using VS:
\begin{displaymath}
  \varphi_{500}^{(i)}=\VS(\exists x_1\dots\exists x_{i-1}\exists
  x_{i+1}\dots\exists x_{11}\varphi_{500}).
\end{displaymath}
Each $\varphi_{500}^{(i)}$ is a univariate quantifier-free formula describing
all possible real choices for $x_i$ for which there exist real choices for all
other variables such that $\varphi_{500}$ holds. CAD can easily decompose the
corresponding one-dimensional spaces. It happens that for each $x_i$ there are
exactly three zero-dimensional cells $a_i$, $b_i$, $c_i\in\R$ where
$\varphi_{500}^{(i)}$ holds. We extract all $a_i$, $b_i$, and $c_i$ as
\emph{real algebraic numbers}, i.e., as the unique root of a univariate defining polynomials with integer coefficients within an isolating interval. By combinatorial arguments it is not hard to see that the
following holds for the set $S_{500}$ of real solutions of $\varphi_{500}$:
\begin{displaymath}
  3\leq|S_{500}|
  \quad\text{and}\quad
  S_{500}\subseteq \prod_{i=1}^{11}\{a_i,b_i,c_i\}.
\end{displaymath}
Notice that at this point we have proven the existence of multiple fixed points of the system for $k_{19}=500$. We
can furthermore compute $S_{500}$ by plugging the $3^{11}$ candidates from the
Cartesian product into $\varphi_{500}$. A straightforward approach requires
arithmetic with real algebraic numbers followed by the determination of the
signs of the results, which is quite inefficient in practice. However, it turns out that interval arithmetic starting with refinements of the isolating intervals of the real algebraic numbers excludes $3^{11}-3$ of the
candidate solutions. Even the three remaining candidates then require no further checking with algebraic numbers since we already know that $|S_{500}|\geq3$. 
The overall CPU time is 71.3
seconds for 11 runs of VS plus 11 runs of CAD, followed by 16 hours for checking
candidates.  Our checking procedure is a file-based
prototype starting a Reduce process for every single of the $3^{11}$ candidates;
there is considerable room for optimization.

For $k_{19}=200$ instead of $500$ all eleven univariate CAD computations yield
unique solutions which can be straightforwardly combined to one unique
solution for the corresponding $\varphi_{200}$. The overall CPU time here is
66.4 seconds for 11 runs of VS plus 11 runs of CAD. Machine float approximations of all
our solutions are given in Table~\ref{TAB:k19fix}.

\begin{table*}[t]
\caption{The unique solution $x^{(200)}$ for $k_{19}=200$ and the three solutions $x^{(500)}_1$, $x^{(500)}_2$, $x^{(500)}_3$ for $k_{19}=500$. 
Note that we have actually computed real algebraic numbers, which are pairs of univariate polynomials and isolated intervals. For convenience we are giving machine float approximations here, which can be made arbitrarily precise.\label{TAB:k19fix}}
\begin{center}
\begin{tabular}{lllll}
  		& $x^{(200)}$ 	& $x^{(500)}_1$	& $x^{(500)}_2$	& $x^{(500)}_3$	\\
\hline  
$x_1$ 	& 90.6512		& 17.6392		& 122.034		& 323.761		\\
$x_2$ 	& 2.67311		& 6.97675		& 14.6721		& 9.49621		\\
$x_3$ 	& 10.4996		& 367.57		& 234.974		& 37.1013		\\
$x_4$ 	& 17.8545		& 36.6772		& 14.5102		& 6.72938		\\
$x_5$ 	& 35.9695		& 5.50874		& 7.16952		& 13.6295		\\
$x_6$ 	& 32.0501		& 12.811		& 35.064		& 43.1428		\\
$x_7$ 	& 0.0954536		& 0.511775		& 0.42579		& 0.127807		\\
$x_8$ 	& 15.5631		& 83.4416		& 69.4223		& 20.8381		\\
$x_9$ 	& 2.39331		& 8.06095		& 7.43877		& 3.21139		\\
$x_{10}$& 0.641001		& 0.25622		& 0.70128		& 0.862856		\\
$x_{11}$& 45.4331		& 2.73253		& 15.2681		& 61.4581		
\end{tabular}
\end{center}
\end{table*}

\subsection{Parametric Analysis for $k_{19}$}
\label{REDLOG:k19}

We next consider the case where $k_{19}$ is left as a free parameter:
\begin{equation}
\varphi_{k_{19}}=\varphi\sigma[100/k_{17},50/k_{18}].
\label{eq:AllButK19}
\end{equation}
Again, we solve for $i\in\{1,\dots,11\}$ eleven QE
problems using VS:
\begin{displaymath}
  \varphi_{k_{19}}^{(i)}=\VS(\exists x_1\dots\exists x_{i-1}\exists
  x_{i+1}\dots\exists x_{11}\varphi_{k_{19}}).
\end{displaymath}
This time each $\varphi_{k_{19}}^{(i)}$ is a bivariate quantifier-free formula
in $k_{19}$ and the corresponding $x_i$.  Hence we must now construct a two-dimensional CAD for each $\varphi_{k_{19}}^{(i)}$. The projection order is
important: we first project $x_i$, then the CAD base phase decomposes the
$k_{19}$-axis, followed by an extension phase that decomposes the $x_i$-space
over the $k_{19}$-cells obtained in the base phase.  This is feasible if we make one limitation: not to extend over zero-dimensional $k_{19}$-cells. In other
words, we accept finitely many blind spots in parameter space, which we can explicitly read off from the CAD so that in the end we know exactly what we are missing.


\begin{sidewaysfigure}[p]
\includegraphics[width=1.1\textwidth]{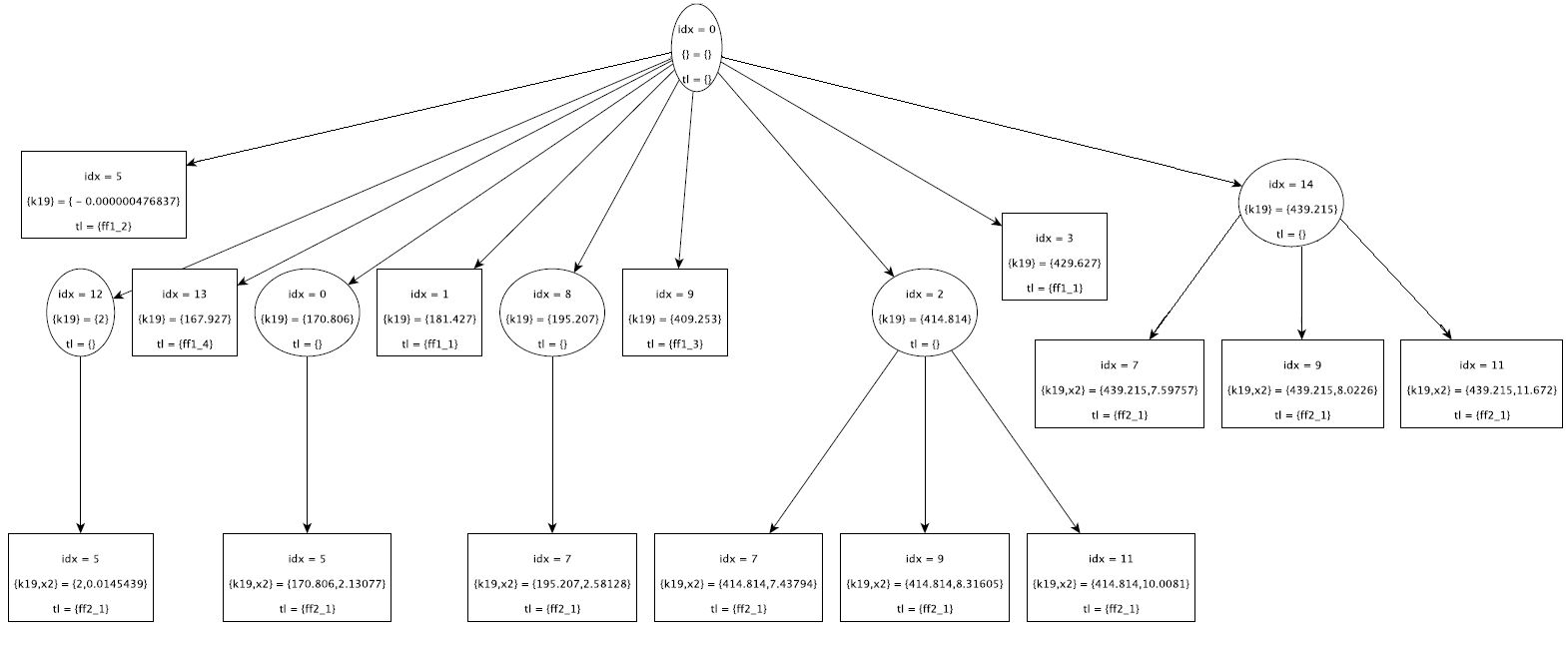}
\caption{The pruned CAD tree for $x_2$. 
Ellipses and rectangles are full-dimensional and zero-dimensional cells, respectively. 
We have removed cells where $k_{19}$ is negative or where the input formula is false.\label{FIG:CAD-x2}}
\end{sidewaysfigure}

Figure~\ref{FIG:CAD-x2} shows our CAD tree for $\varphi_{k_{19}}^{(2)}$. The first layer from the root shows the decomposition of the $k_{19}$-axis. The five zero-dimensional (rectangular) cells are the previously mentioned blind spots, among which the smallest one is not relevant, as it has negative value of $k_{19}$. 
Those zero-dimensional cells also establish the limits of the full dimensional (oval) cells in between. The cylinders over those one-dimensional $k_{19}$-cells each contain either one or three zero-dimensional $x_2$-cells where $\varphi_{k_{19}}^{(2)}$ holds. We have deleted from the tree all $x_2$-cells where $\varphi_{k_{19}}^{(2)}$ does not hold. 

We make two observations, important for a qualitative analysis of our system:
\begin{enumerate}[(i)]
\item For all positive choices of $k_{19}$, extending to infinity, there is at least one positive solution for $x_2$.
\item There is a break point around $k_{19}=409.253$ where the system changes from having a unique solution to exactly three solutions.
\end{enumerate}
Recall that for all floating point numbers given here as approximations we in
fact know exact real algebraic numbers. For instance, the exact break point
is the only real zero in the open interval $(409,410)$ of an irreducible defining polynomial
\begin{equation}
  \label{EQ:definingpol}
  \sum_{i=0}^{10}c_ik_{19}^i\
  \text{with integer coefficients $c_i$ as in \ref{SEC:AppDefPol}}.
\end{equation}

\begin{figure}[p]
  \begin{center}
    \includegraphics[width=\textwidth]{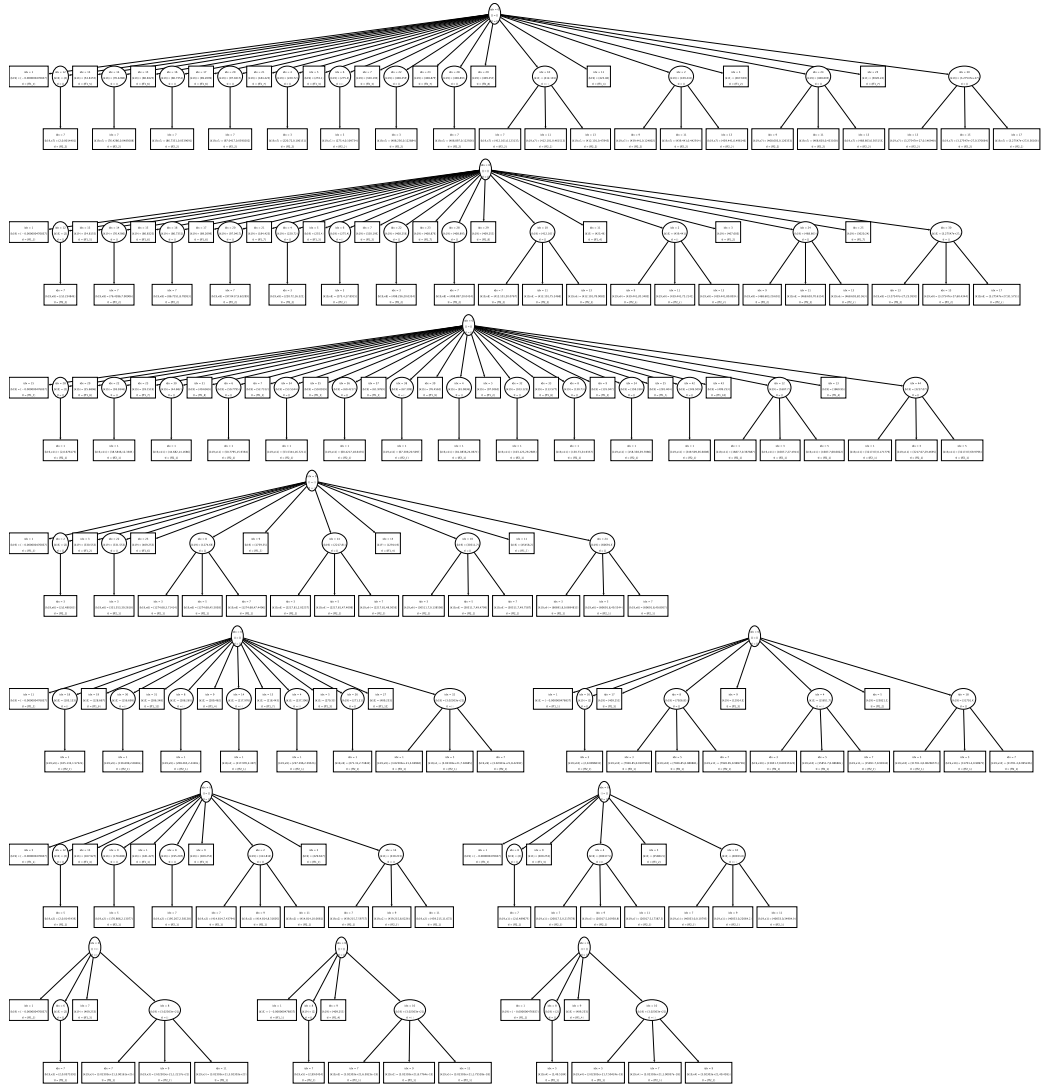}
  \end{center}
\caption{All CAD trees for $\psi_{k_{19}}^{(1)}$, \dots, $\psi_{k_{19}}^{(11)}$. In the second but last row on the left hand side there is the tree for $\psi_{k_{19}}^{(1)}$, which is displayed in detail in Figure~\ref{FIG:CAD-x2}.  Note that in the digital version of this article readers can \emph{zoom into} these trees to see the details (as are visible in the printed version of Figure \ref{FIG:CAD-x2}). \label{FIG:cad-all}}
\end{figure}

Figure \ref{FIG:cad-all} depicts all eleven CAD trees for $\psi_{k_{19}}^{(1)}$,
\dots, $\psi_{k_{19}}^{(11)}$.
They are quite similar to the one just discussed. Even the break point from one to three solutions for $x_i$ is identical for all $i\in\{1,\dots,11\}$ so that we can generalize our observations from earlier:
\begin{enumerate}[(i)]
\item For all positive choices of $k_{19}$, extending to infinity, there is at
  least one positive solution for $(x_1,\dots,x_{11})$.
\item There is a break point $\beta$ around $k_{19}=409.253$ where the system changes its qualitative behaviour. We have exactly given $\beta$ as a real algebraic number in Equation~(\ref{EQ:definingpol}). 
For $k_{19}<\beta$ there is exactly one positive solution for $(x_1,\dots,x_{11})$. For $k_{19}>\beta$ there are at least $3$ and at most $3^{11}$ positive solutions for $(x_1,\dots,x_{11})$.
\end{enumerate}

The overall computation time for our parametric analysis is 4.3~minutes. It is strongly dominated by 2.8 minutes for the computation of one particular CAD tree, for $\varphi_{k_{19}}^{(11)}$. It turns out that the suitable projection order with $x_i$ eliminated first is computationally considerably harder than projecting the other way round. As a preprocessing step we apply CAD-based simplification of the $\varphi_{k_{19}}^{(i)}$ with the opposite, faster, projection order. Here we use QEPCAD-B (v1.69), which performs better than Redlog at simple solution formula construction \citep{Brown2003b}.

\section{Using Triangular Decomposition Tools in the Regular Chains Library for Maple}
\label{SEC:Maple}

In this section we are going to apply triangular decomposition methods, including CAD.  We find that a triangular decomposition can derive solution formulae for many variables in terms of a smaller subset for which we must apply CAD to count solutions.  Throughout this section we are performing computations in Maple 2016, but using an updated version of the Regular Chains Library\footnote{\url{www.regularchains.org}}.  Timings are reported for a Windows 7 64 bit Desktop PC with Intel i5.

\subsection{Parametric Analysis for $k_{19}$}
\label{SUBSEC:Maplek19}

Regular chains are the triangular decompositions of systems of polynomial equations, where triangular means decreasing subsets of variables occurring in each polynomial.  Highly efficient methods for working in complex space have been developed based on these; see the work of \citet{Wang2000} and \citet{ALM99} for a survey.

Recent work by \citet{CDMMXX13} proposes adaptations of these tools to the real analogue: semi-algebraic systems.  They describe two algorithms to decompose any real polynomial system into finitely many regular semi-algebraic systems.  The first, \emph{Real Triangularize} (RT), does so directly while the second, \emph{Lazy Real Triangularize} (LRT), produces the highest (complex) dimension solution component and unevaluated function calls, which if all evaluated would combine to give the full solution.  These algorithms are implemented in the Regular Chains Library for Maple.

We will apply LRT on the quantifier-free formula (\ref{eq:varphi}) evaluated with the parameter estimates for $k_1$, \dots,~$k_{18}$, i.e. the system (\ref{eq:AllButK19}) as studied with Redlog in Section \ref{REDLOG:k19}.

We need to choose a variable ordering: our analysis requires that $k_{19}$ be the indeterminate considered alone.  We place the remaining variables in lexicographical order since the in-built heuristics to make the choice could suggest nothing better.   The solutions must hence contain constraints in $k_{19}$, constraints in ($x_1, k_{19})$, in ($x_2, x_1, k_{19})$ and so on.  

Applying LRT this way produces one solution component and 6 unevaluated function calls in around 15 seconds.  

\subsubsection{The main solution component from LRT}
\label{SSSEC:mainSol}

In the evaluated component: for each of  $x_2$, \dots,~$x_{11}$ there is a single equation which has this as the main variable.  Further, these are all linear in their main variable meaning they can be easily rearranged into the solution formulae given below.
\begin{align}
x_{11} &= -\frac{1}{60}x_{2}^2 
+ \frac{1}{600}(10k_{19} - 10x_{1} - 37x_{3} + 10x_{4} - 2100)x_{2}
-\frac{9}{200}x_{3}^2 \nonumber \\
&\qquad + \frac{1}{600}(-27x_{1} + 27x_{4} + 27k_{19} - 4650)x_{3} - x_{1} + x_{4} + k_{19} - 50
\label{eq:x11Sol} 
\\
x_{10} &= \frac{1}{150}x_{2}(x_{2}+x_{3}-x_{4}-k_{19}+x_{1}+150)
\label{eq:x10Sol}
\\
x_{9} &= \frac{1}{18200}(69x_{3}+182x_{2})(x_{2}+x_{3}-x_{4}-k_{19}+x_{1}+150)
\label{eq:x9Sol}
\\
x_{8} &= \frac{15}{364}(x_{2}+x_{3}-x_{4}-k_{19}+x_{1}+150)x_{3}
\label{eq:x8Sol}
\\
x_{7} &= 50 - \frac{2}{101}x_{4}x_{1} - x_{4}
\label{eq:x7Sol}
\\
x_{6} &= \frac{2}{101}x_{4}x_{1}
\label{eq:x6Sol}
\\
x_{5} &= x_{2} + x_{3} - x_{4} - k_{19} + x_{1} + 150
\label{eq:x5Sol}
\\
x_{4} &= 
2525000/(101x_{2}+1000x_{1}+50500)
\label{eq:x4Sol}
\\
x_{3} &= n_3 / d_3 \quad \mbox{where}
\label{eq:x3Sol}
\\
&n_3 = -101x_{2}^3 - (-101k_{19} + 1101x_{1} + 65650)x_{2}^2 - (1000x_{1}^2 \nonumber \\
&\qquad + (-1000k_{19}+200500)x_{1}  - 50500k_{19}+5050000)x_{2}  + 150000x_{1})\nonumber \\
&d_3 = 101x_{2}^2 + (1000x_{1}+50500)x_{2} \nonumber \\
x_{2} &= n_2 / d_2 \quad \mbox{where} \label{eq:x2Sol} \\
  &n_2 = 30625833064790009548991419920x_{1}^5 \nonumber \\
  &\qquad + (-43795148662369306906962603840k_{19}\nonumber\\
      &\qquad + 37749979225487731805273686504663200)x_{1}^4 \nonumber\\
      &\qquad + (14871210647782462053693235920k_{19}^2 \nonumber\\
      &\qquad - 16963336293692750919154910690672400k_{19} \nonumber\\
      &\qquad + 6815925407229297763234036009365120000)x_{1}^3 \nonumber \cr
      &\qquad + (1538325448222983229930530049200k_{19}^2 \nonumber \\
      &\qquad - 862702164104208291031357996000020000k_{19}\nonumber \cr
      &\qquad + 279241219028720368578809336249748000000)x_{1}^2 \nonumber \\
      &\qquad + (29370341694954648101085099000000k_{19}^2 \nonumber \cr
      &\qquad - 12995812279808313524592161760000000k_{19} \nonumber \\
      &\qquad + 3705960282117523242886769213700000000000)x_{1}\nonumber \cr
      &\qquad - 126235874510278395777369000000000000k_{19}\nonumber \cr
  &d_2 = 232763663752113237974029404420089x_{1}^5 \nonumber\\
  &\qquad + ( - 332853615301041845577671639990228k_{19} \nonumber\cr
      &\qquad + 88646303215205075376308147029677220)x_{1}^4 \nonumber\\
      &\qquad+ (113024761399450186949390623074789k_{19}^2 \nonumber\cr
      &\qquad - 80843908028331498139954527761762740k_{19} \nonumber\\
      &\qquad+ 11682465068391769796632986929072776500)x_{1}^3 \nonumber\cr
      &\qquad + (11455232309649034305597048791479020k_{19}^2  \nonumber\\
      &\qquad- 5547251026060433566640620528023877000k_{19}  \nonumber\cr
      &\qquad + 619147207587597001268026254404647600000)x_{1}^2  \nonumber\\
      &\qquad+ (290245997063001550130198026458525000k_{19}^2  \nonumber\cr
      &\qquad - 141348286758352762323489548674398500000k_{19} \nonumber\cr
      &\qquad + 14547288529581382252587071541494600000000)x_{1} \nonumber\cr
      &\qquad - 1247498501818579946626756931775000000000(k_{19}-100) \nonumber
\end{align}
Note that these solution formula: are guaranteed valid for all positive $k_{19}$ excluding three isolated points which are provided as part of the output from LRT and described below; are triangular, with each $x_k$ is expressed in variables $\{x_i, i<k\}$; and are provided for all but variable $x_1$.

The output of LRT also requires that $x_1$ be both positive and satisfy:
\begin{equation}
\label{eq:x1pol}
f(x_{1}, k_{19}) = \sum_{i=0}^{6} d_ix_{1}^i = 0
\end{equation}
where the coefficients $d_i$ are univariate polynomials in $k_{19}$ of maximum degree $2$ as given in \ref{App:polyF}. 
Hence there are at most six solutions for $x_1$, with the exact number depending on whether solutions of (\ref{eq:x1pol}) are real and positive.

There are four constraints on free parameter $k_{19}$ as given below, one of which is the non-vanishing of the polynomial in Appendix \ref{SEC:AppDefPol} whose root defined the break point found by Redlog in Section \ref{REDLOG:k19}.  Note that the coefficients break over lines within the final constraint.
\begin{align}
  &k_{19}>0\, \label{eqtabk19-c1} \\
  &\land \, \mbox{polynomial in (\ref{EQ:definingpol})}  \neq 0 \, \label{eqtabk19-c2} \\[\lrtskip]
  &\land \, 23197989433419579994929k_{19}^2 
  - 89407400615452409453098800k_{19} \nonumber \\
  &\quad - 4822419303419166525491149190000 \neq 0 \, 
\label{eqtabk19-c3}
  \\[\lrtskip]
  &\land \, 505465566622475867655547880786544637953790406059982726185509k_{19}^4\nonumber \\
  &\quad -  1272578045696439189317856051518387368422217896986836692050 \nonumber\\
  &\quad 5134120k_{19}^3
  + 117551033091520524183124321323141751700303731556 \nonumber \\
  &\quad 2884193657451445400k_{19}^2
  - 281867359883676159811192082978541193 \nonumber\\
  &\quad 600292804324596911878337972560000k_{19}
  - 42434363570215587465 \nonumber \\
  &\quad 668423701563932185051066892741207931879307200000000  \neq 0
\label{eqtabk19-c4}
\end{align}
Evaluating the real roots of the polynomials appearing in the above allows us to conclude that this solution component is valid for all positive values of $k_{19}$ excluding three points.  As with Redlog, Maple can represent these as exact algebraic numbers but for brevity we give float approximations: 
\begin{equation}
\label{eq:blindspots}
409.253, 16473.337, \mbox{ and } 25084.536.
\end{equation}

\begin{sRemark}
\label{rem:Bug1}
In the authors' ISSAC 2017 paper \citep{BioISSAC17} the description of the evaluated solution component ended here.  However, following the publication of that paper a bug was uncovered by one of the authors in the simplifier of the Regular Chains Library when working with a different MAPK model to the one considered presently.  For that example the simplifier was incorrectly discarding certain positivity conditions.  The bug was reported to the Regular Chains developers, and the current version of the simplifier\footnote{\url{http://www.arcnl.org/cchen/software.html}} now excludes all such simplifications.  So presently, the output from LRT includes also the positivity conditions
\[
x_2>0, x_3>0, \dots, x_{11}>0.
\]
Some of these can clearly be removed.  For example, if we know $x_1>0$ and $x_2>0$ then (\ref{eq:x4Sol}) implies $x_4>0$ and this coupled with (\ref{eq:x6Sol}) implies $x_6>0$.  However, it is not trivial to imply all such inequalities, and so any proposed solution in $(k_{19}, x_1)$ should be checked to see if it implies a positive solution in all the remaining variables before being accepted.  This is indeed the case for all solutions described in the ISSAC 2017 paper, and below.
\end{sRemark}

\subsubsection{The unevaluated function calls from LRT}

The main solution component described in Section \ref{SSSEC:mainSol} is not the entire solution to the system.  LRT produced also six unevaluated function calls which if evaluated and combined with the main component would give the full solution.  LRT guarantees that the complex dimension of the solution components from these unevaluated calls is smaller that the main component.  In fact, three of the six unevaluated calls define empty solution sets, with evaluating to discover this instantaneous.

With regards to the other three: we can infer from the arguments to these function calls that each defines the solution at one of the three points in  (\ref{eq:blindspots}) that were excluded from the main component.  I.e. each of these three calls has as an argument the negation of one of the univariate inequations for $k_{19}$ from (\ref{eqtabk19-c2})$-$(\ref{eqtabk19-c4}).  Actually evaluating these solution components is not possible in reasonable time.  Thus, as with Redlog in Section \ref{SEC:Redlog}, we proceed accepting a small number of blind spots.  

The output of LRT has quickly given us the structure of the solution space valid at all but three isolated values of $k_{19}$.  However, it does not identify where the number of real solutions change.  Note that  although the break point identified in Section \ref{SEC:Redlog} has been rediscovered in (\ref{eq:blindspots}), there is not yet any information gathered by Maple from which we can infer its significance.  We also note that there seems to be no significance for our application of the other two isolated points in (\ref{eq:blindspots}).  

\subsubsection{Counting solutions with CAD}

To finish the analysis we need to decompose $(x_1, k_{19})$-space according to the real roots of $f(x_1, k_{19})$; and also $x_1$ and $k_{19}$ since the constraints $x_1>0$ and $k_{19}>0$  were specified separately in the output.
CAD is ideally suited for this task. We apply the Regular Chains based implementation in Maple first described by \citet{CMXY09}.  A CAD for $f(x_{1}, k_{19})$, with the ordering chosen so that the $k_{19}$-axis is the one decomposed, divides the plane into 135 cells in a few seconds.  This CAD decomposes the $k_{19}$ axis into 11 cells, i.e. identifying five points, which approximate to: 
\[
-379.993, \, -87.776, \, 0, \, 409.253, \mbox{ and } 25084.536.
\] 
We give these as floats for brevity but exact algebraic numbers are available\footnote{See the Research Data Statement at the end of the paper to access them.}.

On the cell where $0<k_{19}<409.253$, the cylinder above in the $(x_1, k_{19})$-plane is divided into 11 cells: three of which cover $x_1>0$ (two 2d sectors and a 1d section).  We see that $f(x_1, k_{19})$ is zero on the section but not the sectors.  This can be inferred by testing a sample point of the section (the invariance properties of the CAD mean that the signs of the input at this point are representative for the whole cell.  In fact, with the CAD implementation we use the cells comes with a semi-algebraic description which for this section is the statement that $f(x_1, k_{19})=0$ (along with the bounds on $k_{19}$).

We can perform a similar analysis on the two cells for $409.253<k_{19}< 25084.536$ and $25084.536<k_{19}<\infty$.  In each case the cylinders above are divided into 15 cells, seven of which cover $x_1>0$, with the three sections satisfying $f(x_1, k_{19})=0$.  

So we can conclude that:  (a) if $0<k_{19}<409.253$ then $f(x_1, k_{19})$ has a single positive real solution; and (b) if $k_{19} \in (409.253, \infty) \setminus \{25084.536\}$ then $f(x_1, k_{19})$ has three positive real solutions.  We cannot conclude with certainty what happens at the points $409.253$ and $25084.536$.

At the end of this analysis we have rediscovered the break point identified in Section \ref{SEC:Redlog} where the system moves from a single positive real solution to three.  We also have explicit solutions valid for all except three isolated $k_{19}$ values.  To obtain an actual numerical solution we need only:
select the $k_{19}$ value of interest (call it $\hat{k}_{19}$); 
perform univariate root isolation on $f(x_1, \hat{k}_{19})$,  noting we know in advance how many to expect based on $\hat{k}_{19}$; 
then for each $x_1$ solution substitute recursively into equations (\ref{eq:x11Sol})$-$(\ref{eq:x2Sol}), starting with (\ref{eq:x2Sol}) and working up, substituting the new variable solution from each formula into the next.  
The solutions in Table \ref{TAB:k19fix} may be easily rediscovered this way, for example.  

We note that, as discussed in Software Remark \ref{rem:Bug1}, we have ensured that for each cell all the positive solutions in $x_1$ provided by the sample point do indeed lead to positive solutions for all other variables via the back substitution process.

\subsection{Repeating for Other Choices}
\label{SUBSEC:MapleOther}

We have repeated the approach described in Section \ref{SUBSEC:Maplek19} for different choices of free parameter and different choices of fixed parameter values.  For example:
\begin{itemize}
\item With $k_{17}$ set to $95$ instead of $100$ we find that the break point between 1 and 3 real positive solutions moves to $k_{19} = 369.917$.  With $k_{17}$ set to $105$ it moves to $k_{19} = 450.077$.
\item Allowing $k_{17}$ to be free and fixing $k_{19} = 200$ we find that there is only ever one positive real solution.
\item Allowing $k_{17}$ to be free and fixing $k_{19} = 500$ we find the number of positive real solutions moving from 1 to 3 to 1 breaking at $k_{17} = 85.988$ and $k_{17} = 110.869$.  
\item Similarly, allowing $k_{18}$ to be free and fixing $k_{19} = 200$ we find there is only ever one positive real solution; but fixing $k_{19}=500$ instead we find 3 real solutions between $k_{18} = 44.434$ and $58.329$ and 1 otherwise.  
\end{itemize}
This hints that there is a shape approximating a paraboloid within ($k_{17}$, $k_{18}$, $k_{19}$)-space within which bistability may occur; with bistability available for any $k_{17}$ and $k_{18}$ value but bounded from below in the $k_{19}$ coordinate.

We note that these conclusions are, as with the one described in detail, valid at all but a handful of isolated values of the free parameter.  

\section{A Graph Theory Guided Parametric Gaussian Elimination Preprocessing Method}
\label{SEC:Preproc}

As described above, the complexity of polynomial systems obtained with steady-state approximations of biological models is comparatively high for the application of symbolic methods, particularly in reference to the dimension (number of indeterminates).  The two studies described in Sections \ref{SEC:Redlog} and \ref{SEC:Maple} both used tools to effectively reduce the problem dimension before applying the costly CAD method.  

More generally, it is highly relevant for the the success of general polynomial systems methods if we can first identify and exploit particular structural properties of the input.   
Here, the MAPK models have remarkably low total degrees with many linear monomials after some substitutions for rate constants.  For example, the final equation of $\overline{(\ref{EQ:thesystem})}$ suggests a simple polynomial expression for $x_{11}$ in terms of the remaining variables of the system.
This promoted the idea of pre-processing MAPK input with essentially Gaussian elimination: in the sense of solving single suitable equations with respect to some variable and substituting the corresponding solution into the system.

\subsection{Parametric Gaussian Elimination}

Generalizing this idea to situations where linear variables have parametric coefficients in the other variables requires, in general, a parametric variant of Gaussian elimination, which replaces the input system with a finite case distinction with respect to the vanishing of certain coefficients and one reduced system for each case.  Further, for our problem the positivity conditions establish a further apparent obstacle, because we are formally not dealing with a parametric system of linear equations but with a parametric linear programming problem.

The theory of real quantifier elimination by virtual substitution tells us that it is sufficient for the inequality constraints to play a passive role in the sense that their polynomials do not contribute to the elimination set $E$ discussed in Section~\ref{SE:vsintro}. This key idea occurred first for the linear case in Theorem 3.11 of the work by \citet{LoosWeispfenning:93a}; while the current state-of-the-art is described in the thesis of \citet{Kosta:16a}.  The crucial observation is that our entire formula is (and remains during the considered elimination) a single Gauss Prime Constituent in the sense of \citep[Section 3.1.1]{Kosta:16a}.  Further, for the considered MAPK model, it turns out that those positivity assumptions on the variables are actually strong enough to guarantee the non-vanishing of all relevant coefficients, so case-distinctions are never necessary!  
We do not claim such an approach will always be so lucky, but it may be this result generalises for the MAPK hierarchy.  It was the case also for the second larger MAPK model we describe in Section \ref{SEC:Mod28}.

\subsection{An Optimal Strategy}
  
Parametric Gaussian elimination can increase the degrees of variables in the parametric coefficient, in particular destroying their linearity and suitability to be used for further reductions.  For example, solving the last equation of $\overline{(\ref{EQ:thesystem})}$ and substituting into the first equation would destroy any linearity present in that first equation.

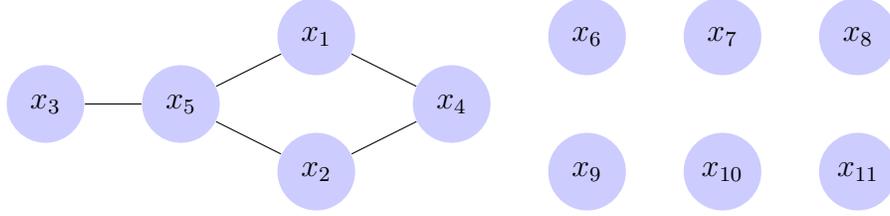
\begin{figure}[t]
  \centering
  \begin{tikzpicture}[scale=.9,auto=left,every
    node/.style={circle,fill=blue!20,minimum size = 2.5em}]
    \node (n3) at (1,20) {$x_3$}; \node (n5) at (3,20) {$x_5$}; \node (n1) at
    (5,21) {$x_1$}; \node (n2) at (5,19) {$x_2$}; \node (n4) at (7,20)
    {$x_4$};

    \node (n6) at (9,21) {$x_6$}; \node (n7) at (11,21) {$x_7$}; \node (n8) at
    (13,21) {$x_8$}; \node (n9) at (9,19) {$x_9$}; \node (n10) at (11,19)
    {$x_{10}$}; \node (n11) at (13,19) {$x_{11}$};

    \foreach \from/\to in {n3/n5,n1/n5,n1/n4,n2/n4,n2/n5} \draw (\from) --
    (\to);
  \end{tikzpicture}
  \caption{The graph for $\overline{(\ref{EQ:thesystem})}$ is loosely connected. Its minimum vertex cover $\{x_4,x_5\}$ is small. All other variables form a maximum independent set, which can be eliminated with linear methods.\label{fig:vc1}}
\end{figure}

The natural question is whether there is an optimal strategy to Gauss-eliminate a maximal number of variables?  This has been answered positively only recently by \citet{Grigoriev2015}: draw a graph, where vertices are variables and edges indicate multiplication between variables within some monomial. Then one can Gauss-eliminate a \emph{maximum independent set}, which is the complement of a \emph{minimum vertex cover}. Figure~\ref{fig:vc1} shows that graph for $\overline{(\ref{EQ:thesystem})}$, where $\{x_4,x_5\}$ is a minimal vertex cover, and all other variables can be linearly eliminated. 

Recall that minimum vertex cover is one of 21 classical NP-complete problems described by \citet{Karp:72}. However, our instances considered here and instances to be expected from other biological models are so small that the use of existing approximation algorithms \citep{Grandoni2008} appears unnecessary. We have used real quantifier elimination, which did not consume measurable CPU time; alternatively one could use integer linear programming or SAT-solving.

It is a most remarkable fact that a significant number of biological models in the databases have that property of loosely connected variables. This phenomenon resembles the well-known \emph{community structure} of propositional satisfiability problems, which has been identified as one of the key structural reasons for the impressive success of state-of-the-art CDCL-based SAT solvers by \citet{girvan2002community}.

\subsection{Reduced System for Model 26}

We conclude this section with the reduced system computed with an implementation of this pre-processing in Redlog~\citep{DolzmannSturm:97a}. From (\ref{eq:3ParaTarski}) we obtain 
\begin{align}
\overline{\psi} &= x_{5} >0 \,\land\, 
x_{4} > 0 \,\land\, 
k_{19} > 0 \,\land\, 
k_{18}> 0 \,\land\, 
k_{17} > 0 \nonumber \\
&\quad \land  1062444 k_{18} x_{4}^{2} x_{5} + 23478000 k_{18} x_{4}^{2} + 1153450 k_{18}
  x_{4} x_{5}^{2} + 2967000 k_{18} x_{4} x_{5} 
\nonumber \\
  &\qquad + 638825 k_{18} x_{5}^{3} + 49944500 k_{18} x_{5}^{2} - 5934 k_{19}
  x_{4}^{2} x_{5} - 989000 k_{19} x_{4} x_{5}^{2}
\nonumber \\
  &\qquad - 1062444 x_{4}^{3} x_{5} -
  23478000 x_{4}^{3} - 1153450
  x_{4}^{2} x_{5}^{2}- 2967000 x_{4}^{2} x_{5}
\nonumber \\
  &\qquad - 638825 x_{4} x_{5}^{3} -
  49944500 x_{4} x_{5}^{2} = 0
\nonumber \\
&\quad \land 1062444 k_{17} x_{4}^{2} x_{5} + 23478000 k_{17} x_{4}^{2} + 1153450
  k_{17}x_{4} x_{5}^{2} + 2967000 k_{17} x_{4} x_{5}
\nonumber \\
  &\qquad+ 638825 k_{17}
  x_{5}^{3}
  + 49944500 k_{17} x_{5}^{2} - 1056510 k_{19} x_{4}^{2} x_{5} - 164450 k_{19}
  x_{4} x_{5}^{2}
\nonumber \\
  &\qquad- 638825 k_{19} x_{5}^{3} - 1062444 x_{4}^{2} x_{5}^{2} - 23478000 x_{4}^{2}
  x_{5} - 1153450 x_{4} x_{5}^{3}
\nonumber \\
  &\qquad - 2967000 x_{4} x_{5}^{2} -
  638825 x_{5}^{4} - 49944500 x_{5}^{3} = 0.
\label{eq:M26Red}
\end{align}
We now have a system of just  two equalities in 5 indeterminates together with positivity conditions on those indeterminates.  Notice that no complicated positivity constraints come into existence from this method. All corresponding substitution results are entailed by the other constraints, which is implicitly discovered by using the standard simplifier of \citet{DolzmannSturm:97c} during preprocessing.

Note that, with $\psi$ defined in (\ref{eq:3ParaTarski}), we have a formal equivalence here, from the theory of quantifier elimination via virtual substitution:
\[
\exists x_1 \exists x_2 \dots \exists x_{11} \, \psi = \exists x_4 \exists x_5 \, \overline{\psi}.
\]
So if we can determine the region of parameter space where solutions to $\overline{\psi}$ exist we are guaranteed to also find solutions to $\psi$ there.  However, our problem concerns not just the existence of solutions but the number, and so on the surface this may seems inadequate.  However, because the only technology used in this reduction is linear substitution we can also conclude that the number of solutions found for $\overline{\psi}$ will lead to the same number of solution of $\psi$.

Hence it is sufficient to study $\overline{\psi}$.  This pre-processing allows us to derive solutions with two free-parameters in the next section.  We also give some indication of the performance improvements of various methods offered by the pre-processing later in Section \ref{SEC:vsNumerical}.

\section{Combined Approach for a Solution over 2-parameter space}
\label{SEC:New2Para}

In this section we describe a new derivation of a solution to the real algebraic problem with two free parameters, produced after the publication of the authors' ISSAC 2017 and CASC 2017 conference papers \citep{BioISSAC17, BioCASC17}.  The progress is made by combining ideas from all three of the preceding sections.
We describe in detail below but broadly we:
start with the reduced system from the pre-processing of Section \ref{SEC:Preproc} with two free-parameters;
apply the LRT method of Section \ref{SEC:Maple} to reduce the problem by an indeterminate;
build part of a CAD, an idea used in Section \ref{SEC:Redlog}, sufficient to identify the regions of parameter space of interest.  Timings are reported for the same hardware and software as Section \ref{SEC:Maple}.

\subsection{Applying LRT and Preparing for CAD}
\label{SUBSEC:2ParaLRT}

We start with the reduced system (\ref{eq:M26Red}) derived in Section \ref{SEC:Preproc} above.  We set $k_{18}$ to 50 and leave $k_{17}$ and $k_{19}$ free.  Hence we seek the regions of the $(k_{17}, k_{19})$-plane where there exist multiple solutions.  

We first run the LRT algorithm introduced in Section \ref{SEC:Maple}, using variable ordering $(x_4, x_5, k_{17}, k_{19})$.  We needed the parameters to come after the variables so we work over the parameter space, but within the pairs the orders could have been reversed.
In around 5 seconds LRT outputs one solution component and 4 unevaluated function calls.  

The evaluated component consists of the four positivity conditions from the input and the two equations, which may be seen in \ref{SEC:AppSol} where they are labelled (\ref{eq:withX4}) and (\ref{eq:withoutX4}).
Of course these equations are triangular: (\ref{eq:withX4}) involves $\{x_4, x_5, k_{17}, k_{19}\}$ while (\ref{eq:withoutX4}) does not depend on $x_4$.
Note that (\ref{eq:withX4}) is linear in $x_4$ and so we can easily rearrange to give a solution formula for $x_4$ in terms of $(x_5, k_{17}, k_{19})$.  
(\ref{eq:withoutX4}) is of degree 6 in $x_5$ but of course not all its solutions need be real and positive.  If we can determine where (\ref{eq:withoutX4}) has multiple positive real solutions then all that remains is to back substitute and to get real solutions for the other variables and check these are also positive.  We will determine this using CAD.

Before that, we examine the 4 unevaluated functions calls from LRT: two instantaneously evaluate to empty solution sets while the other two cannot be evaluated in reasonable time.  We infer from the arguments to the function calls that the latter two define solutions on the graphs of two polynomials in $(k_{17}, k_{19})$-space.  These two polynomials may be found in \ref{SEC:AppExcl}.  The smaller is degree 5 in $k_{17}$ and degree 4 in $k_{19}$ (total degree 5 overall) and the larger degree 14 in $k_{17}$ and degree 10 in $k_{19}$ (total degree 14 overall)\footnote{As described later in Section \ref{SUBSEC:Conjecture} the boundary of the multistationarity region is actually defined by part of the graph of one of these polynomials, although there is no reason to conclude that at this stage of the analysis.}.

We proceed on the understanding that any results are valid everywhere in $(k_{17}, k_{19})$-space except on these graphs.  We may compare this to Sections \ref{REDLOG:k19} and \ref{SUBSEC:Maplek19} which accepted a finite number of isolated blind spots in a one-dimensional parameter space.

\subsection{Solution via an Open CAD}
\label{SUBSEC:2ParaCAD}

A CAD sign-invariant for the polynomial defining $(\ref{eq:withoutX4})$ (and $x_5, k_{17}, k_{19}$ to allow for positivity checks) would be sufficient.  However, the size of the polynomial puts this beyond CAD currently.  Instead, we proceed as follows:
\begin{description}
\item[Step 1:] Calculate the projection set for CAD input consisting of polynomial defining (\ref{eq:withoutX4}) and polynomial $x_5$ (to allow for positivity check).  
\end{description}
This is a set of 19 polynomials in $(k_{17}, k_{19})$ the greatest of which has degree 34, and so it is not reasonable to print them all here.
\begin{description}
\item[Step 2:] Build an Open CAD of $(k_{17}, k_{19})$-space for these polynomials, along with polynomials $k_{17}$ and $k_{19}$ (to allow for positivity checks).  
\end{description}
An \emph{Open CAD} means the full dimensional cells only.  The boundaries may be determined by algebraic numbers but because we do not lift over the boundaries there no costly algebraic number calculations.  The idea has been much discussed by \citet{McCallum1993, Strzebonski2000, WBDE14}, and other names used for it include generic CAD and 1-layered Sub-CAD.  It was partly applied by the approach in Redlog in Section \ref{SEC:Redlog}.  It is sufficient to solve problems which are only in strict inequalities, but of course, that is not the case here.  By making this restriction we are accepting that our solutions and conclusions are not necessarily valid on cell boundaries: a finite number of curve segments in the $(k_{17}, k_{19})$-plane.  However, we have already made such an acceptance, in the use of LRT above.  

We perform the above steps with the ProjectionCAD package of \citet{EWBD14} in Maple\footnote{\url{http://computing.coventry.ac.uk/~mengland/ProjectionCAD.html}} in 17 seconds.  The resulting CAD has 533 cells.

\begin{description}
\item[Step 3:] Identify those cells in the upper quadrant of the $(k_{17}, k_{19})$-plane. 
\end{description}
We only care about solutions in this upper quadrant.  We can easily identify 139 such cells by querying sample points (note that no cell can straddle the boundary of the quadrant since the CAD produced was also produced sign-invariant for $k_{17}$ and $k_{19}$ as polynomials).  Since in Step 1 we ensured that this CAD was built for the projection of the polynomial defining (\ref{eq:withoutX4}) we may conclude that for this polynomial we can work at a sample point of the cell but draw conclusions for the whole cell, as we do next.

\begin{description}
\item[Step 4:] Identify the number of positive real roots the polynomial defining (\ref{eq:withoutX4}) has over each of these cells.
\end{description}

We do this by substituting for the sample point and applying Maple's default real root isolation algorithm.  We identify 35 of the 139 cells where there are three positive real roots for $x_5$, with the other 104 all having one.

\begin{description}
\item[Step 5:] Check that these solutions provide a positive solution for $x_4$  via back substitution into (\ref{eq:withX4}).
\end{description}
We first checked that the 104 cells with one positive real solution for $x_5$ all lead to one positive real solution for $x_4$ as expected.  We then analyse the 35 cells and each of their three positive real solutions for $x_5$ in turn.  For 28 of these cells each solution gives a corresponding positive real solution for $x_4$.  For the other 7 cells, only one of the three solutions does, so these join the other 104 as representing the parameter space with one solution.

\begin{figure}[p]
  \begin{center}
    \includegraphics[width=.48\textwidth]{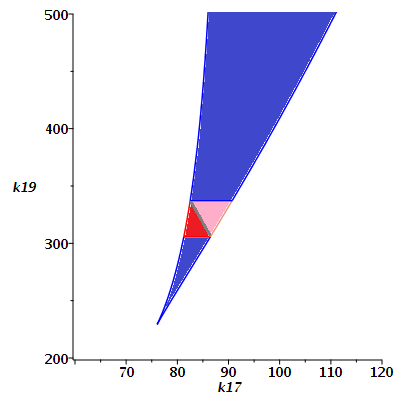}
    \includegraphics[width=.48\textwidth]{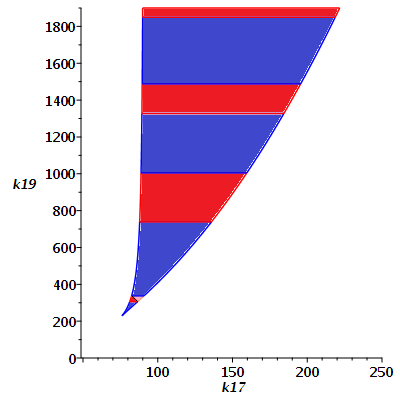}
  \end{center}
\caption{Visualisations of the Open CAD cells describing the multistationarity region derived in Section \ref{SEC:New2Para} for smaller values of $k_{17}$ and $k_{19}$. \label{fig:MSRegion1}}
\end{figure}

\begin{figure}[p]
  \begin{center}
    \includegraphics[width=.48\textwidth]{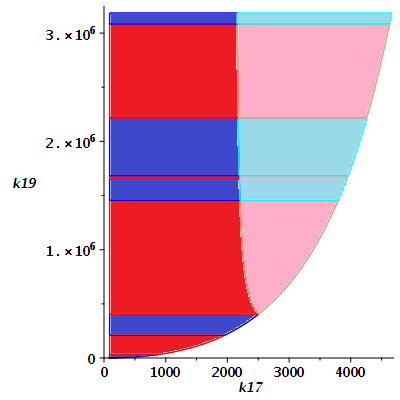}
    \includegraphics[width=.48\textwidth]{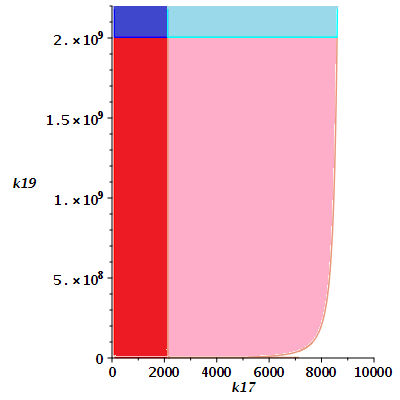}
  \end{center}
\caption{Visualisations of the Open CAD cells describing the multistationarity region derived in Section \ref{SEC:New2Para} for larger values of $k_{17}$ and $k_{19}$. \label{fig:MSRegion2}}
\end{figure}

The semi-algebraic descriptions of these 28 cells provide the exact description of the regions in $(k_{17}, k_{19})$-space where multistationarity can occur. We use these descriptions to produce the 4 plots of the multistationarity region in Figures \ref{fig:MSRegion1} and \ref{fig:MSRegion2}.  The 4 images are all produced from the data in the 28 cells, but with different plotting regions.  In each case, the coloured regions represent the cells with multistationarity, with the only purpose of the different colours to show the separation of the cells\footnote{Because we produced an Open CAD above we cannot formally conclude what happens on these cell boundaries.}.  

The left plot in Figure \ref{fig:MSRegion1} is for the original range of $k_{19}$ values considered and has the region of multistationarity described by 4 full dimensional CAD cells.  The right plot shows that this region grows as $k_{19}$ increases: at this range 9 cells are in view including the 4 from the left plot which are at the bottom of the region.  

The left plot of Figure \ref{fig:MSRegion2} expands the ranges considerably.  There are 24 cells in view of the range but the original 9 described above are now too small to see.  The right plot of Figure \ref{fig:MSRegion2} expands the range further to include all 28 cells; with all 24 from the previous image now too small to see.  In this final image the two cells at the top actually extend infinitely in the $k_{19}$ direction while always being bounded on both sides in the $k_{17}$ direction.

\section{Stability of Fixed Points}
\label{SEC:Stability}

The work described in Section \ref{SEC:Redlog}$-$\ref{SEC:New2Para} was dedicated to identifying where multiple fixed points occur.  This alone does not prove multistationarity as we must also check the stability properties of these fixed points.

We may use the three linear conservation constraint equations (\ref{EQ:claws}) to eliminate $x_1$, $x_7$, and $x_{11}$ from system (\ref{EQ:thesystem}) and symbolically compute the Jacobian $\tilde{J}$ of the obtained reduced system.  We can then numerically compute the eigenvalues of $\tilde{J}$ for the instances arising from the substitution of the parameter values and the different positive fixed points for the variables. 

We have used the float approximations for the unique solution $x^{(200)}$ with $k_{19}=200$ and the three solutions $x^{(500)}_1$, \dots,~$x^{(500)}_3$ for $k_{19}=500$ in Table \ref{TAB:k19fix}.   For the single positive fixed point $x^{(200)}$ the Jacobian $\tilde{J}(x^{(200)})$ has eigenvalues with negative real part only and hence can be shown to be stable. For $k_{19}=500$ one of the three positive fixed points $x^{(500)}_2$ can be shown to be unstable, as $\tilde{J}(x_2^{(500)})$ has one eigenvalue with positive real part; the other seven had negative real parts.  In contrast $x_1^{(500)}$ and $x_3^{(500)}$ can be shown to be stable. Hence for $k_{19}=500$ the system is indeed bistable.

A verification of the stability of the fixed points using exact real algebraic numbers by the well-known Routh--Hurwitz criterion is possible algorithmically \citep{HongLiskaSteinberg97}, but seems to be out of range of current methods for this example.
Notice that in other studies on multistationarity of signaling pathways, such as those of \citet{Conradi2008} and \citet{Gross2016}, the question of stability has also been left to one side.

\section{Another MAPK Model}
\label{SEC:Mod28}

We describe a second MAPK model, which we will use alongside the first from Section \ref{SEC:Problem} in the remaining sections, to broaden the conclusions drawn.

\subsection{MAPK Bio-Model 28}
\label{SUBSEC:Mod28}

The system with number 28 in the BioModels Database is given by the following set of differential equations.  This model is the distributive fully random kinetics version of the models proposed by \citet{Markevich2004}. Hereafter we refer to it as Model 28.  Again, we have renamed the species to $x_1, \ldots, x_{16}$ and the rate constants to $k_1, \ldots, k_{27}$ to facilitate reading:
\begin{eqnarray}
\difft{x}_{1} & = & k_2 x_9 + k_8 x_{10} + k_{21} x_{15} + k_{26} x_{16} \nonumber \\
& & \qquad 
- k_1 x_1 x_5 - k_7 x_1 x_5 - k_{22} x_1 x_6 - k_{27} x_1 x_6  \nonumber \\
\difft{x}_{2} &= & k_3 x_9 + k_5 x_7 + k_{24} x_{12} - k_4 x_2 x_5 - k_{23} x_2 x_6 \nonumber \\
\difft{x}_{3} & = & k_9 x_{10} + k_{11} x_8 + k_{16} x_{13} + k_{19} x_{14} - k_{10} x_3 x_5 - k_{17} x_3 x_6 - k_{18} x_3 x_6 \nonumber \\
\difft{x}_{4} &=&  k_6 x_7 + k_{12} x_8 + k_{14} x_{11} - k_{13} x_4 x_6 \nonumber \\
\difft{x}_{5} &= & k_2 x_9 + k_3 x_9 + k_5 x_7 + k_6 x_7 + k_8 x_{10} + k_9 x_{10} + k_{11} x_8 + k_{12} x_8 -\nonumber \\
                  &    & \quad k_1 x_1 x_5 - k_4 x_2 x_5 - k_7 x_1 x_5 - k_{10} x_3 x_5  \nonumber \\
\difft{x}_{6} & = & k_{14} x_{11} + k_{16} x_{13} + k_{19} x_{14} + k_{21} x_{15} + k_{24} x_{12} + k_{26} x_{16} - \nonumber \\
                &    & \quad  k_{13} x_4 x_6 - k_{17} x_3 x_6 - k_{18} x_3 x_6 - k_{22} x_1 x_6 - k_{23} x_2 x_6 - k_{27} x_1 x_6  \nonumber \\
\difft{x}_{7} & = & k_4 x_2 x_5 - k_6 x_7 - k_5 x_7  \nonumber \\
\difft{x}_{8} &=  & k_{10} x_3 x_5 - k_{12} x_8 - k_{11} x_8  \nonumber \\
\difft{x}_{9} &= & k_1 x_1 x_5 - k_3 x_9 - k_2 x_9  \nonumber \\
\difft{x}_{10} &= &k_7 x_1 x_5 - k_9 x_{10} - k_8 x_{10}  \nonumber \\
\difft{x}_{11} &=& k_{13} x_4 x_6 - k_{15} x_{11} - k_{14} x_{11}  \nonumber \\
\difft{x}_{12}& = &  k_{23} x_2 x_6 - k_{25} x_{12} - k_{24} x_{12}  \nonumber \\
\difft{x}_{13} &= &k_{15} x_{11} - k_{16} x_{13} + k_{17} x_3 x_6  \nonumber \\
\difft{x}_{14}& = & k_{18} x_3 x_6 - k_{20} x_{14} - k_{19} x_{14}  \nonumber \\
\difft{x}_{15}& = &k_{20} x_{14} - k_{21} x_{15} + k_{22} x_1 x_6  \nonumber \\
\difft{x}_{16}& = &k_{25} x_{12} - k_{26} x_{16} + k_{27} x_1 x_6  \label{EQ:thesystem28} 
\end{eqnarray}
We denote by $\overline{(\ref{EQ:thesystem28})}$ the system formed by replacing all left hand sides of (\ref{EQ:thesystem28}) by $0$.

The estimates of the rate constants given in the  BioModels Database are:
\begin{align}
  k_{1} &= 0.005,&
  k_{2} &= 1,&
  k_{3} &= 1.08,&
  k_{4} &= 0.025,\nonumber\\
  k_{5} &= 1,&
  k_{6} &= 0.007,&
  k_{7} &= 0.05,&
  k_{8} &= 1,\nonumber\\
  k_{9} &= 0.008,&
  k_{10} &= 0.005,&
  k_{11} &= 1,&
  k_{12} &= 0.45,\nonumber\\
  k_{13} &= 0.045,&
  k_{14} &= 1,&
  k_{15} &= 0.092,&
  k_{16} &= 1,&\nonumber\\
  k_{17} &= 0.01,&
  k_{18} &= 0.01,&
  k_{19} &= 1,&
  k_{20} &= 0.5,&\nonumber\\
  k_{21} &= 0.086,&
  k_{22} &= 0.0011,&
  k_{23} &= 0.01,&
  k_{24} &= 1,&\nonumber\\
  k_{25} &= 0.47,&
  k_{26} &= 0.14,&
  k_{27} &= 0.0018.  \label{EQ:rcestimates28}
\end{align}
Again, using the left-null space of the stoichiometric matrix under positive conditions as a conservation constraint \citep{Famili2003} we obtain the following three linear conservation constraints:
\begin{eqnarray}
 x_6 + x_{11} + x_{12} + x_{13}+ x_{14} + x_{15} + x_{16}   &=&  k_{28}, \nonumber \\
 x_5 + x_7 + x_8 + x_9 + x_{10} &=& k_{29} ,\nonumber \\
 x_1 + x_2 + x_3 + x_4 + x_7 + x_8 + x_9 + x_{10} + x_{11} + {} & &\nonumber \\
 \quad  x_{12} + x_{13} + x_{14} + x_{15} + x_{16}  &=& k_{30}, 
\label{EQ:claws28}
\end{eqnarray}
where $k_{28}$, $k_{29}$, $k_{30}$ are new constants.  Meaningful values for these three are harder to obtain than the constants in (\ref{EQ:rcestimates}).  The following are some realistic value estimates:
\begin{align}
  k_{28} &= 100,&  k_{29} &= 180,& k_{30}&=800.
  \label{EQ:clestimates2}
\end{align}

Ideally we would treat all three symbolically and identify multistationarity within the $(k_{28}, k_{29}, k_{30})$ parameter space.

\subsection{Preprocessing}
\label{SUBSEC:Mod28PreProc}

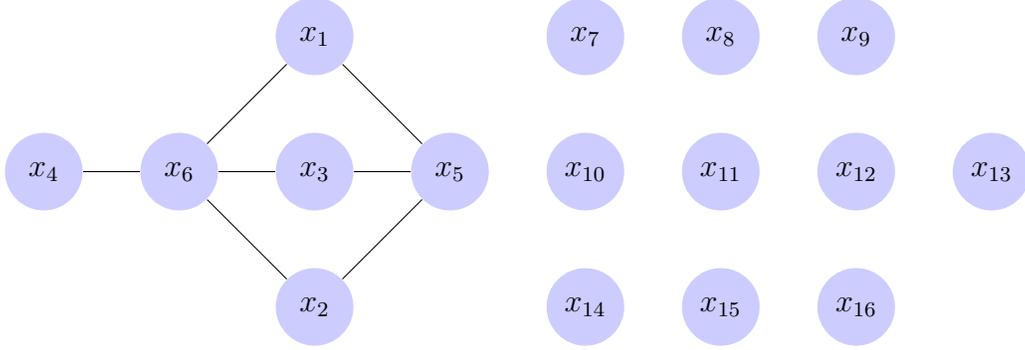
\begin{figure}[t]
  \centering
  \begin{tikzpicture}[scale=.9,auto=left,every
    node/.style={circle,fill=blue!20,minimum size = 2.5em}]
    \node (n4) at (1,20) {$x_4$}; 
    \node (n6) at (3,20) {$x_6$}; 
    \node (n1) at (5,22) {$x_1$}; 
    \node (n3) at (5,20) {$x_3$}; 
    \node (n2) at (5,18) {$x_2$}; 
    \node (n5) at (7,20) {$x_5$};

    \node (n7) at (9,22) {$x_7$}; 
    \node (n8) at (11,22) {$x_8$}; 
    \node (n9) at (13,22) {$x_9$}; 
    \node (n10) at (9,20) {$x_{10}$}; 
    \node (n11) at (11,20) {$x_{11}$}; 
    \node (n12) at (13,20) {$x_{12}$};
    \node (n13) at (15,20) {$x_{13}$}; 
    \node (n14) at (9,18) {$x_{14}$}; 
    \node (n15) at (11,18) {$x_{15}$};
    \node (n16) at (13,18) {$x_{16}$};        

    \foreach \from/\to in {n4/n6, n6/n1,n6/n2,n1/n5,n2/n5,n6/n3,n3/n5} \draw (\from) --
    (\to);
  \end{tikzpicture}
  \caption{The graph for $\overline{(\ref{EQ:thesystem28})}$ produced according to the techniques setout in Section \ref{SEC:Preproc}.
Despite being a larger system the minimum vertex cover $\{x_5,x_6\}$ is still small. All other variables form a maximum independent set, which can be eliminated with linear methods.\label{fig:vc2}}
\end{figure}

We may apply the preprocessing procedure outlined in Section \ref{SEC:Preproc} to $\overline{(\ref{EQ:thesystem28})}$ and the positivity constrains similarly to as described in Section \ref{SEC:Preproc} for Model 26.  
The connection graph is given in Figure \ref{fig:vc2} showing that $\{x_5,x_6\}$ as a minimum vertex cover.  We obtain the simplified system:
\begin{eqnarray*}
  3796549898085 k_{29} x_{5}^{3} x_{6} + 71063292573000 k_{29} x_{5}^{3} &&\\
  + 106615407090630 k_{29} x_{5}^{2} x_{6}^{2}
  {}+ 479383905861000 k_{29} x_{5}^{2} x_{6} &&\\
  + 299076127852260 k_{29} x_{5} x_{6}^{3}
  {}+ 3505609439955600 k_{29} x_{5} x_{6}^{2} &&\\
   + 91244417457024 k_{29} x_{6}^{4}
  {}+ 3557586742819200 k_{29}x_{6}^{3} &&\\
  - 598701732300 k_{30} x_{5}^{3} x_{6}
  {} - 83232870778950 k_{30} x_{5}^{2} x_{6}^{2}&&\\
  - 185019487578700 k_{30} x_{5}x_{6}^{3}
   - 3796549898085 x_{5}^{4} x_{6}&&\\
   - 71063292573000 x_{5}^{4}
  - 106615407090630 x_{5}^{3}
  x_{6}^{2}&&\\
  {} - 479383905861000 x_{5}^{3} x_{6} - 299076127852260 x_{5}^{2}
  x_{6}^{3}&&\\
  - 3505609439955600 x_{5}^{2} x_{6}^{2}
  {}- 91244417457024 x_{5}x_{6}^{4}&&\\
   - 3557586742819200 x_{5} x_{6}^{3} = 0,&&
\end{eqnarray*}
\begin{eqnarray*}
  3796549898085 k_{28} x_{5}^{3} x_{6} + 71063292573000 k_{28} x_{5}^{3} &&\\
  + 106615407090630 k_{28} x_{5}^{2} x_{6}^{2}
  {}+ 479383905861000 k_{28} x_{5}^{2}x_{6} &&\\
  + 299076127852260 k_{28} x_{5} x_{6}^{3}
  {}+ 3505609439955600 k_{28} x_{5} x_{6}^{2} &&\\
  + 91244417457024 k_{28} x_{6}^{4}
  {}+ 3557586742819200 k_{28} x_{6}^{3}&&\\
  - 3197848165785 k_{30} x_{5}^{3} x_{6}
  {} - 23382536311680 k_{30}x_{5}^{2} x_{6}^{2} &&\\
  - 114056640273560 k_{30} x_{5} x_{6}^{3}
  {}- 91244417457024k_{30} x_{6}^{4} &&\\
  - 3796549898085 x_{5}^{3} x_{6}^{2} - 71063292573000x_{5}^{3} x_{6}&&\\
  {}- 106615407090630 x_{5}^{2} x_{6}^{3}
  - 479383905861000x_{5}^{2} x_{6}^{2} &&\\
  - 299076127852260 x_{5} x_{6}^{4} - 3505609439955600 x_{5}&&\\
  x_{6}^{3} - 91244417457024 x_{6}^{5} - 3557586742819200 x_{6}^{4} = 0.&&
\end{eqnarray*}
along with positivity constraints $x_{6} >0$, $x_{5} > 0$, $k_{30} > 0$, $k_{29}> 0$, and $k_{28} > 0$.

\section{Grid Sampling: Symbolic vs Numeric}
\label{SEC:vsNumerical}

In this section we summarise work that was first presented in CASC 2017 \citep{BioCASC17} which compared the use of symbolic and numeric techniques to identify multistationary regions via grid sampling.  

\subsection{Algorithms and Software}
\label{SUBSEC:GSSetup}

In this section we will use Symbolic Grid Sampling: so we have results only for a set of numerical sample points, but each sample point will undergo a symbolic computation. The result will still be an approximate identification of the region, since the sampling will be finite, but the results at those sample points will be guaranteed free of numerical errors.  The symbolic computations follow exactly the strategy introduced in Section \ref{SEC:Maple} except each sample point will set all parameters (rather than leaving one free) meaning a simpler symbolic computation than in Section \ref{SEC:Maple} performed multiple times.  In particular, with no free parameters the Lazy variant of Real Triangularization (LRT) used in Section \ref{SEC:Maple} gives the full solution (no laziness) as we would get from Real Triangularization (RT) and so we just use the latter.  

We will compare this symbolic grid sampling with a fully numerical gird sampling approach using the homotopy solver Bertini developed by \citet{BHSW06}, in its standard configuration to compute complex roots.  Alternatives to Bertini include PHCpack by \citet{Verschelde2011} and the Numerical Algebraic Geometry package for Macaulay2 by \citet{Leykin2011}.  Reasons for choosing Bertini include that it is the most cited homotopy solver for the past 8 years and that it allows adaptive and very high-precision arithmetic (whereas PHCpack only allows double-double)\footnote{We note that a recent development for Bertini published after this article was in press could be applicable to this problem:  Paramotopy by \citet{BBN18} allows for parallelism and computation reuse, well suited for such grid sampling.}.
We parsed the output of Bertini using Python, and determined numerically which of the complex roots are real and positive using a threshold of $10^{-6}$ for positivity. 

Bertini computations (v1.5.1) were carried out on a Linux  64 bit Desktop PC
with Intel i7. Maple computations (v2016 with April 2017 Regular Chains)
were carried out on a Windows 7 64 bit Desktop PC with Intel i5.  

\begin{sRemark}
\label{rem:Bug2}
For the reduced system of Model 28 Bertini (incorrectly) could not find any roots, not even complex ones, for any of the parameter settings. The situation did not change when going from adaptive precision to a very high fixed precision. However, we have not attempted more sophisticated techniques like providing user homotopies. It seems a bug in Bertini has been triggered by this problem instance.  It has been reported to the developers.
\end{sRemark}

\subsection{Sample Ranges and Plots}
\label{SUBSEC:GSRanges}

For Model 26 we will use a sampling range for $k_{19}$ from 200 to 1000 by 50; for $k_{17}$ from 80 to 200 by 10; and for $k_{18}$ from 5 to 75 by 5.  

For Model 28 we will use a sampling range for $k_{30}$ from 100 to 1600 by 100; for $k_{28}$ from 40 to 160 by 10; and for $k_{29}$ from 120 to 240 by 10.

We produce 2d plots in each case with the third parameter fixed to its values indicated in (\ref{EQ:clestimates}) and (\ref{EQ:clestimates2}).
In those plots we will colour sample points according to the number of fixed points observed: yellow discs indicate one fixed point and blue boxes three.  Diamonds indicate numerical errors where zero (red) or two (green) fixed states were identified. 

\subsection{Results and Comparison}
\label{SUBSEC:NvsSResults}

The plots produced by the grid sampling are presented in Figures \ref{FIG:Bertini-Sys26-Original}$-$\ref{FIG:Maple-Sys28}; and the time taken to produce them is summarised in Table \ref{TAB:SysTime}.

\begin{table}[b]
  \caption{Timing data (in seconds) of the grid samplings described in
    Section~\ref{SEC:vsNumerical}. Numerical is using Bertini and Symbolic the Regular Chains Library for Maple.
    \label{TAB:SysTime}}
\begin{center}    
  \begin{tabular}{lr@{\qquad}rrrrrrr}
     & \multicolumn{1}{c@{\,}}{\textbf{Numerical}} & \multicolumn{4}{c}{\textbf{Symbolic}}\\
Model    & Mean   
                      & Mean   & Median & StdDev & Maximum \\ \hline
    26 -- Original 	& 2.4~    
                      & 0.568  & 0.530  & 0.107 & 0.905   \\
    26 -- Reduced  	& 0.85            
                      & 0.053  & 0.047  & 0.036 & 0.343   \\
    28 -- Original 	& 16.57                  
                      & 42.430 & 40.529 & 8.632 & 84.116 \\
    28 -- Reduced  	& $\bot$              
                      & 0.485  & 0.468  & 0.119 & 0.796       
  \end{tabular}
\end{center}
\end{table}

\subsubsection{Comparison of models}

Model 28 forms a larger real algebraic problem than Model 26, 16 variables and equations rather than 11, so it unsurprising that it takes longer to perform computations.

Regarding the symbolic computations: Model 28 requires an actual CAD of a plane to be produced for each sample point while Model 26 only real root isolation (decomposition of a line).  This was the case regardless of whether the original or reduced system was used as the starting point, since the RT preprocessing also reduced the number of variables that needed analysis by CAD.  
We note that even with the reduced system it was still beneficial to pre-process CAD with RT: the average time per sample point with pre-processing (and including time taken to pre-process) was 0.485 seconds while without it was 3.577 seconds.  It is not clear if this is because of a genuine simplification or because the CAD algorithm from the Regular Chains Library that we used it particularly tuned for triangular systems.

\subsubsection{Effects of the pre-processing in Section \ref{SEC:Preproc}}

\begin{figure}[p]
  \centering
  \includegraphics[width=0.49\textwidth]{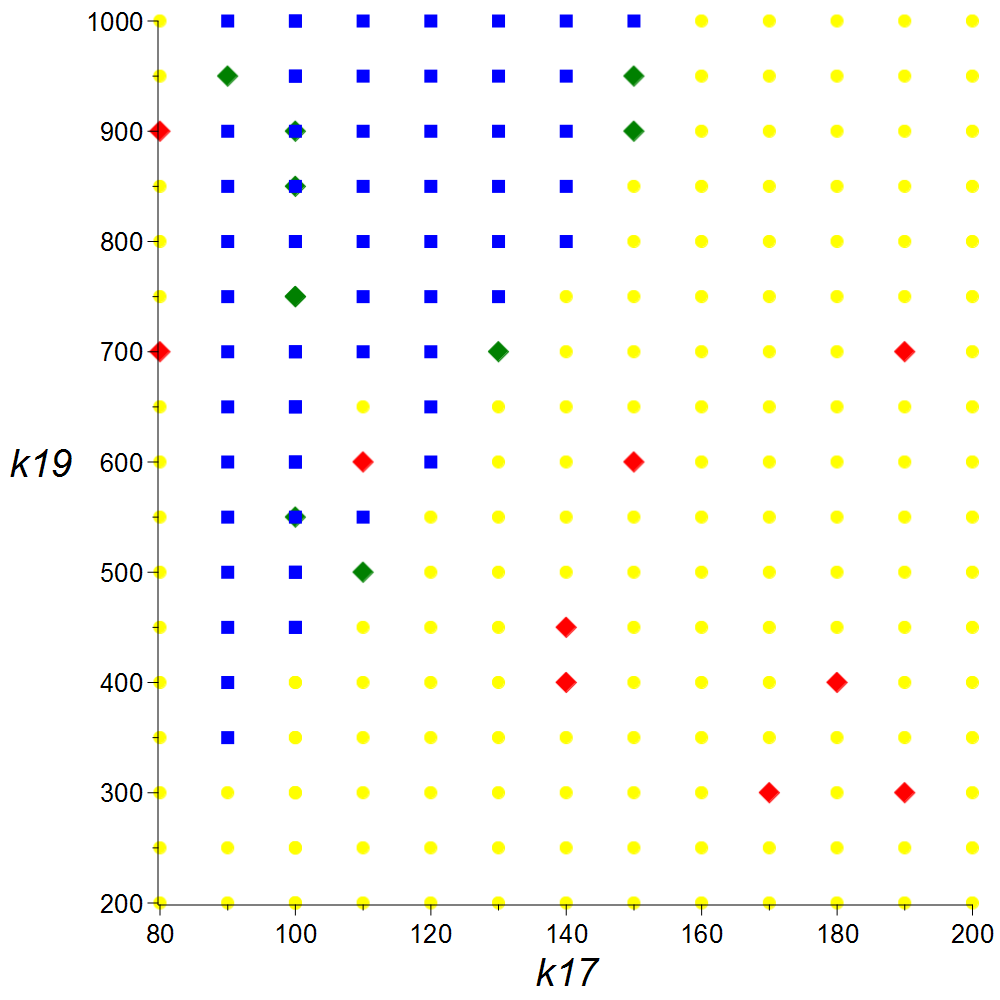}
  \includegraphics[width=0.49\textwidth]{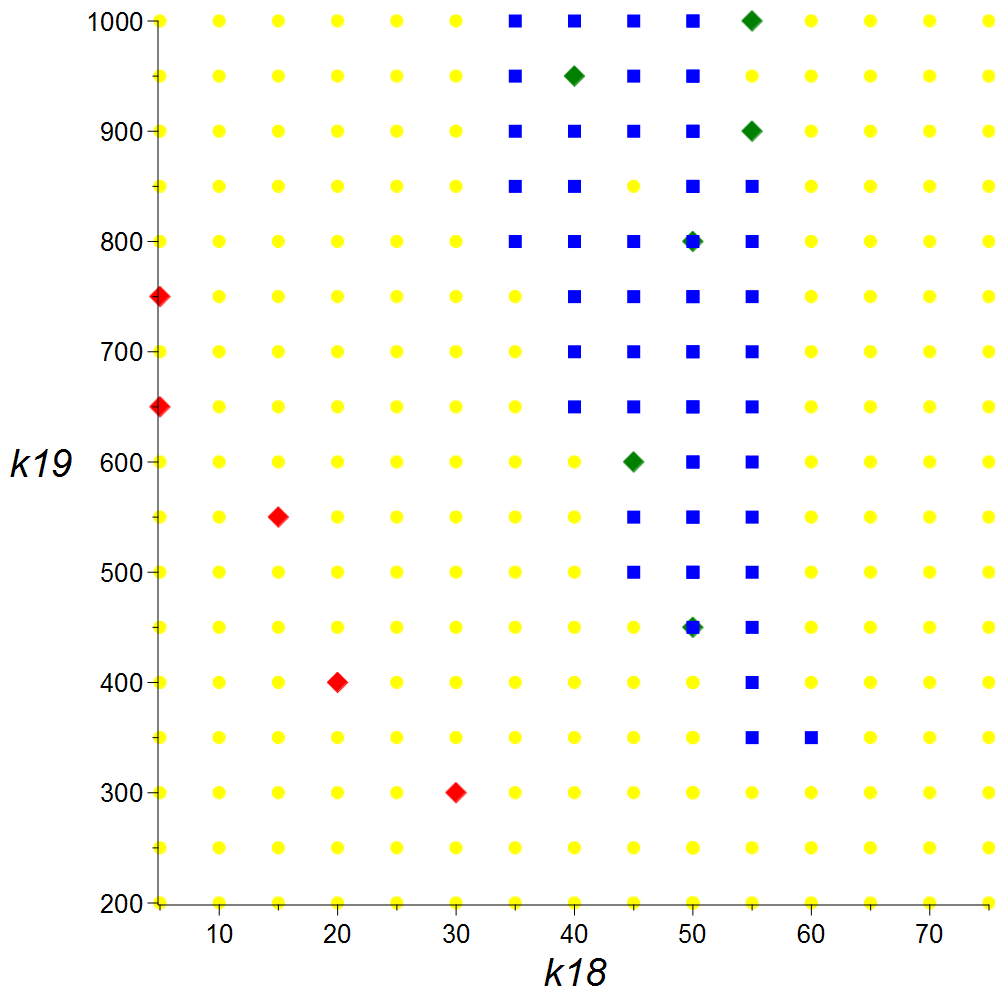}
  \caption{Plots illustrating the result of Bertini's grid sampling on the original version of Model 26.\label{FIG:Bertini-Sys26-Original}}
\end{figure}

\begin{figure}[p]
  \centering
  \includegraphics[width=0.49\textwidth]{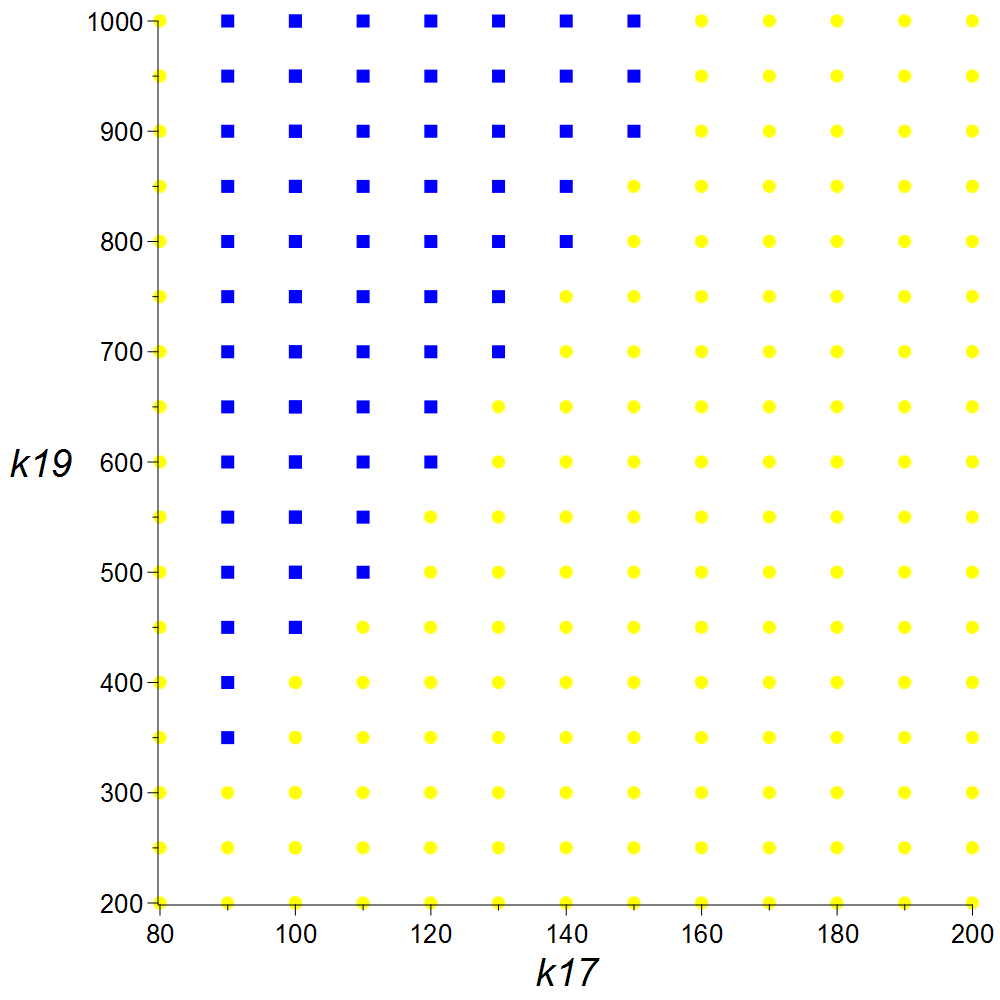}
  \includegraphics[width=0.49\textwidth]{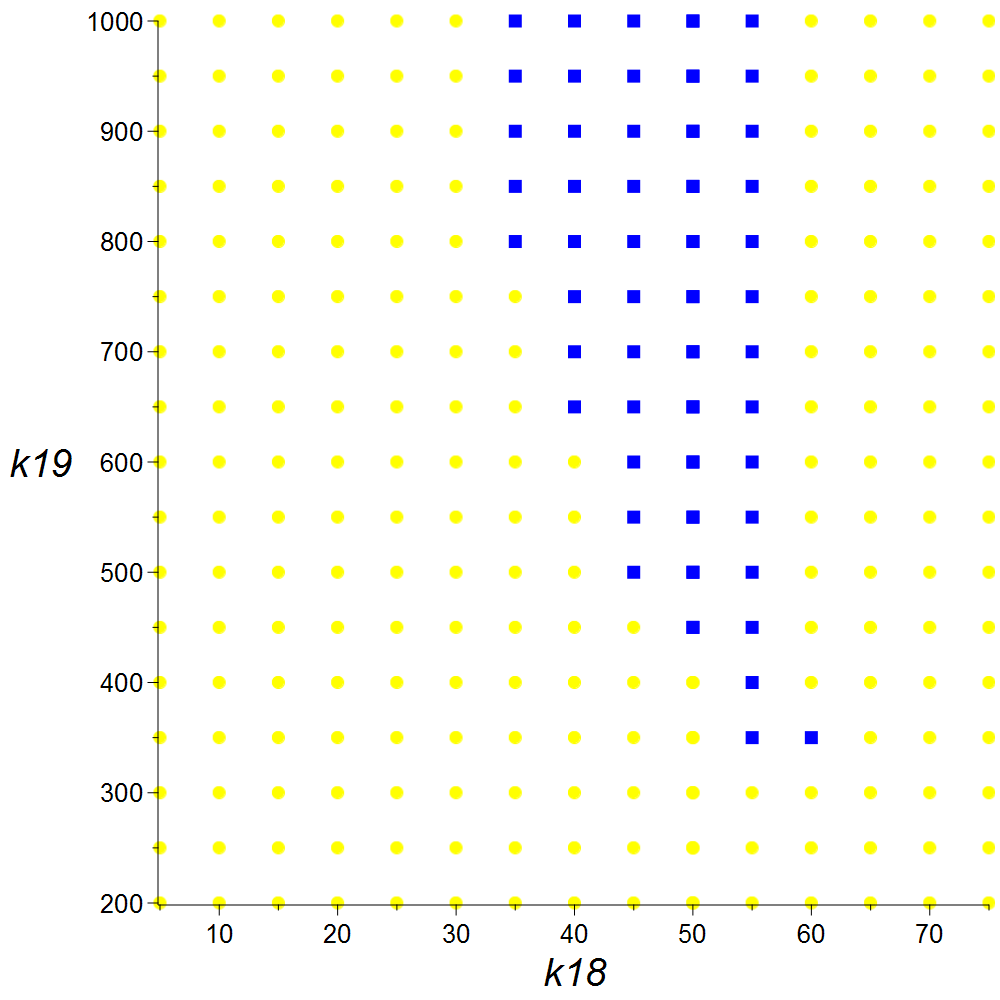}
  \caption{Plots illustrating the result of Bertini's numerical grid sampling on the reduced version of Model 26.  These are also identical to those plots produced by Maple's symbolic grid sampling of Model 26 (both original and reduced versions).  \label{FIG:Good-Sys26}}
\end{figure}

Figure~\ref{FIG:Bertini-Sys26-Original} and Figure~\ref{FIG:Good-Sys26} both refer to Model 26.  The latter is produced by Maple's symbolic calculations and so guaranteed free of numerical error.  
The former, Figure~\ref{FIG:Bertini-Sys26-Original}, represents the output of Bertini on the original system.  We see that there are numerous numerical errors present: the rouge red and green diamonds in Figure~\ref{FIG:Bertini-Sys26-Original}.  
We find that when computing with the reduced system rather than the original system Bertini was able to to avoid all these errors, producing the same plots as Maple in Figure~\ref{FIG:Good-Sys26}.

\begin{figure}[p]
  \centering
  \includegraphics[width=0.49\textwidth]{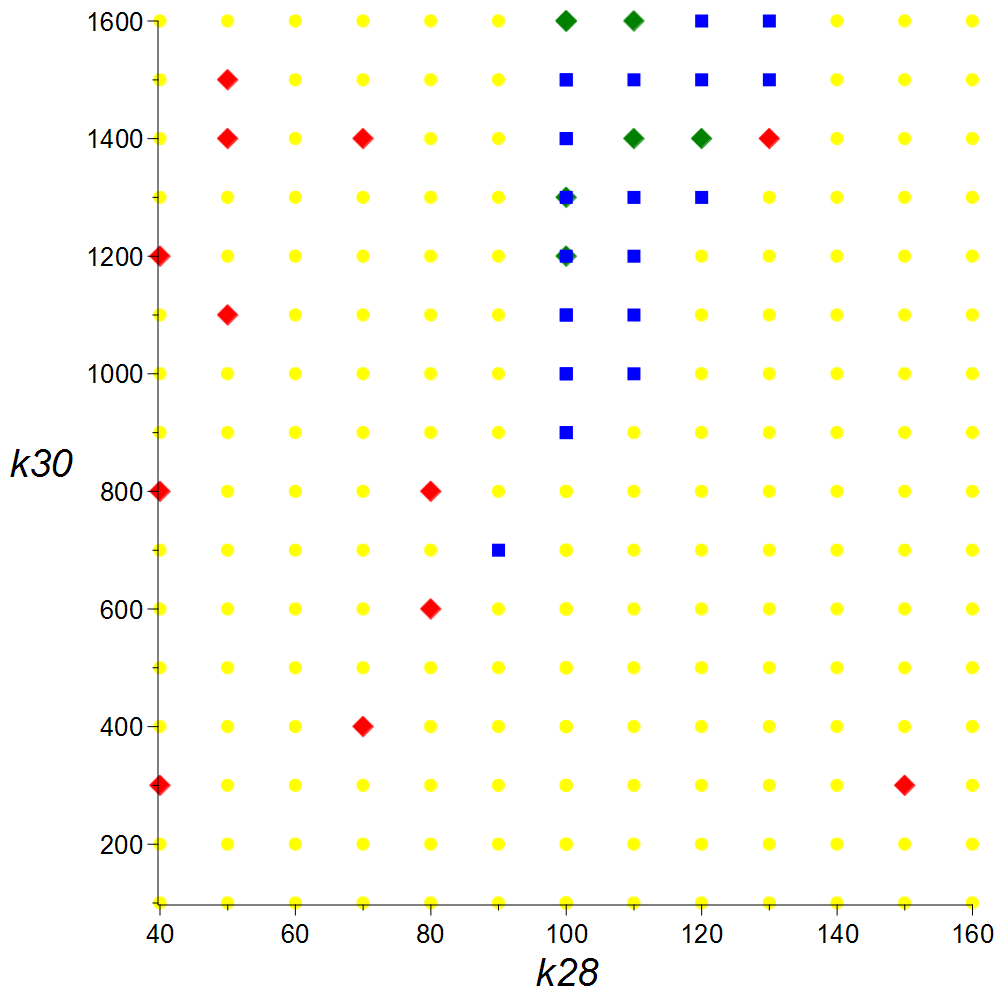}
  \includegraphics[width=0.49\textwidth]{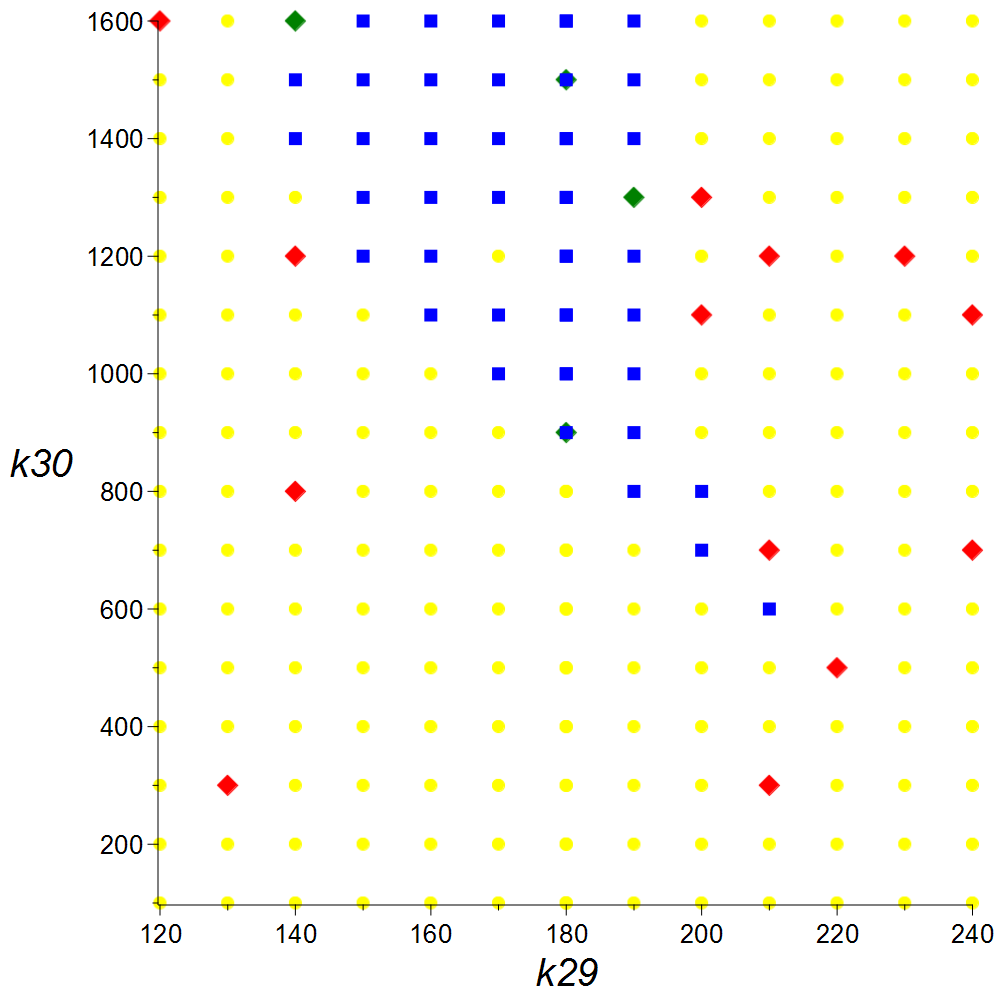}
  \caption{Plots illustrating the result of Bertini's grid sampling on the original version of Model 28.  
  \label{FIG:Bertini-Sys28-Original}}
\end{figure}

\begin{figure}[p]
  \centering
  \includegraphics[width=0.49\textwidth]{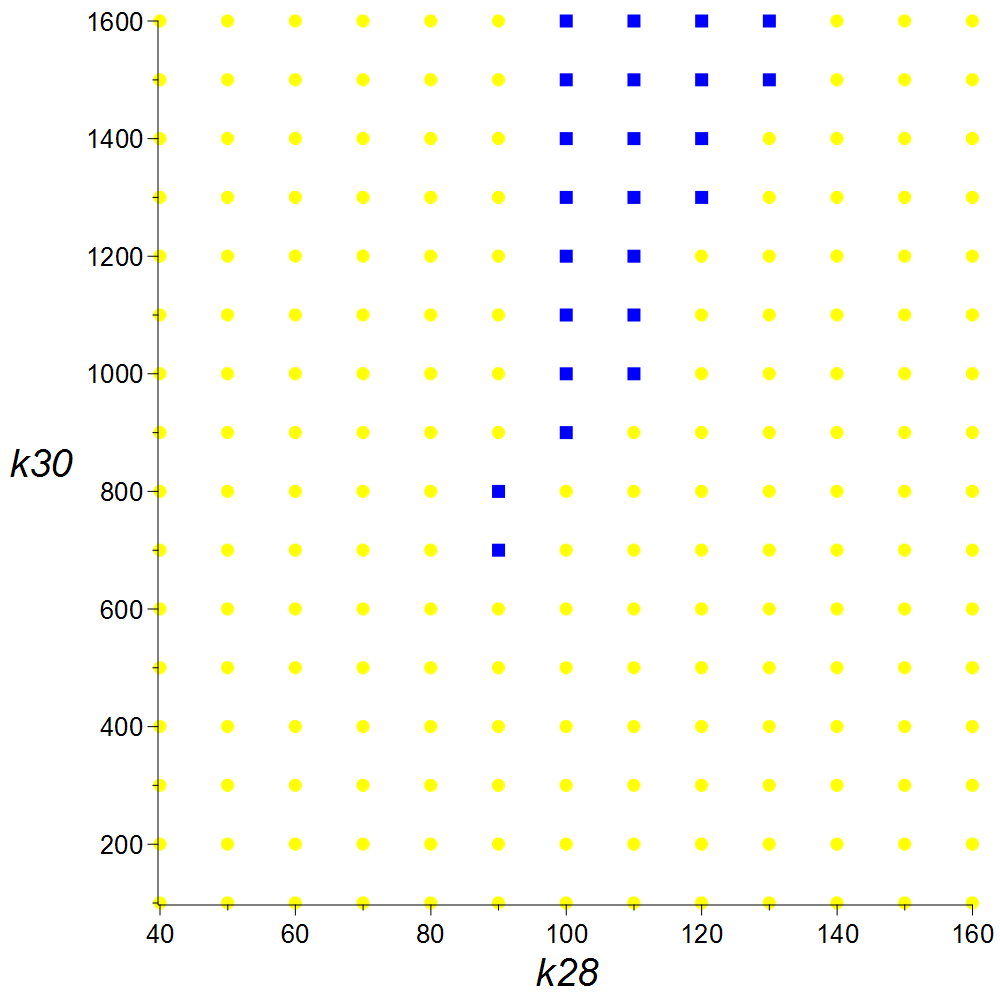}
  \includegraphics[width=0.49\textwidth]{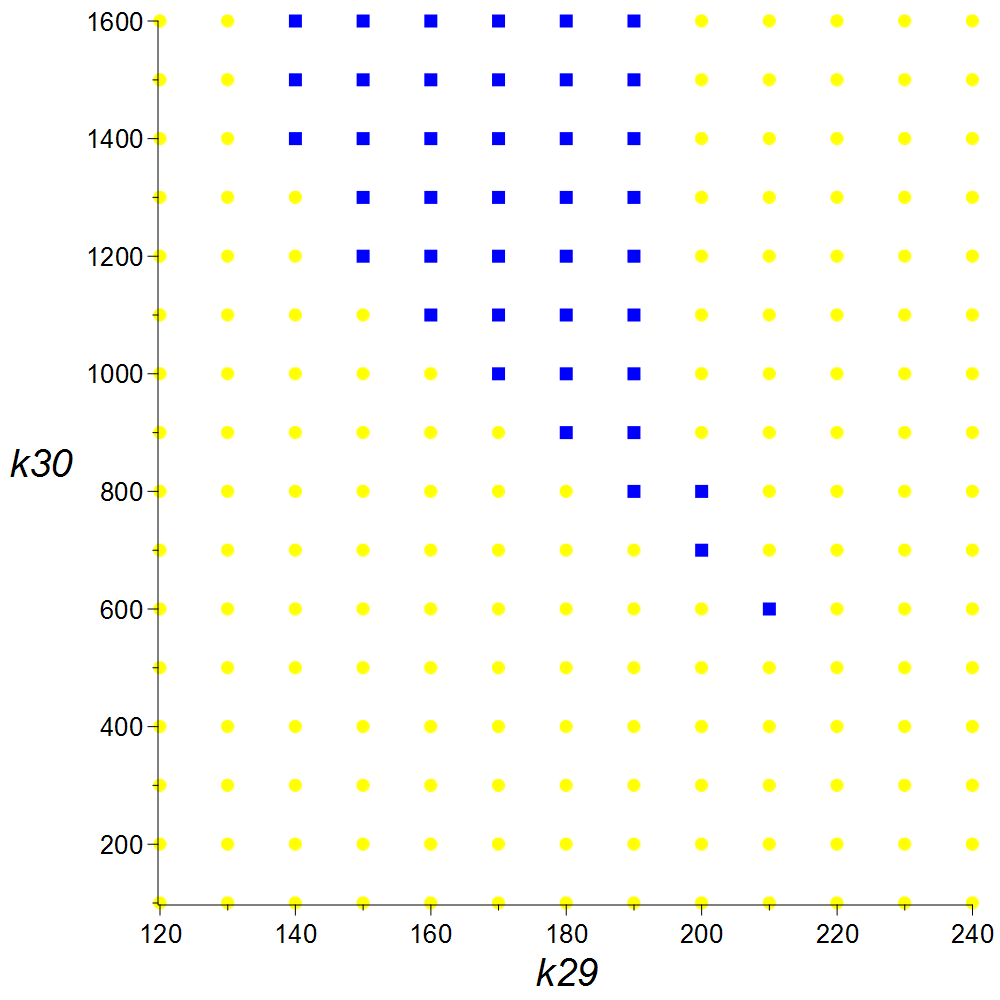}
  \caption{Plots illustrating the result of Maple's symbolic grid sampling on Model 28 (both original and reduced versions).
  \label{FIG:Maple-Sys28}}
\end{figure}

With Model 28 we see similar numerical errors from Bertini in Figure~\ref{FIG:Bertini-Sys28-Original} when compared with Maple in Figure~\ref{FIG:Maple-Sys28}.  However, in the case of Model 28 the reduction led to catastrophic effects for Bertini: built-in heuristics quickly (and incorrectly) concluded that there are no zero dimensional solutions for the system, and when switching to a positive dimensional run also no solutions could be found.

From the timing data in Table \ref{TAB:SysTime} we see that both Bertini and Maple benefited from the reduced system: For Model 26 Bertini took a third of the original time while Maple took a tenth of the original.  For Model 28 the speed-up enjoyed by the symbolic method from the pre-processing was even greater: almost 100 fold!

\subsubsection{Symbolic vs Numerical}

As described above, we have observed numerous numerical errors when using Bertini which may avoided with the symbolic computations of Maple.  However, they can also be avoided (at least for Model 26) by using the pre-processing technique described in Section \ref{SEC:Preproc}.

However, and surprisingly, for Model 26 the symbolic methods were actually quicker than the numerical ones.  The symbolic methods used are well known for their doubly exponential computational complexity (in the number of variables) so it is not necessary surprising that as the system size increases the results of the comparison would change.  For Model 28 we have the expected outcome of the numerical calculations being quicker.
 
We can see some other statistical data for the timings in Maple: the standard deviation for the timings is fairly modest but in each row there are large outliers and so the median is always a little less than the mean average.

\subsection{Higher Sampling Rates}

\begin{figure}[h]
  \centering
  \includegraphics[width=0.49\textwidth]{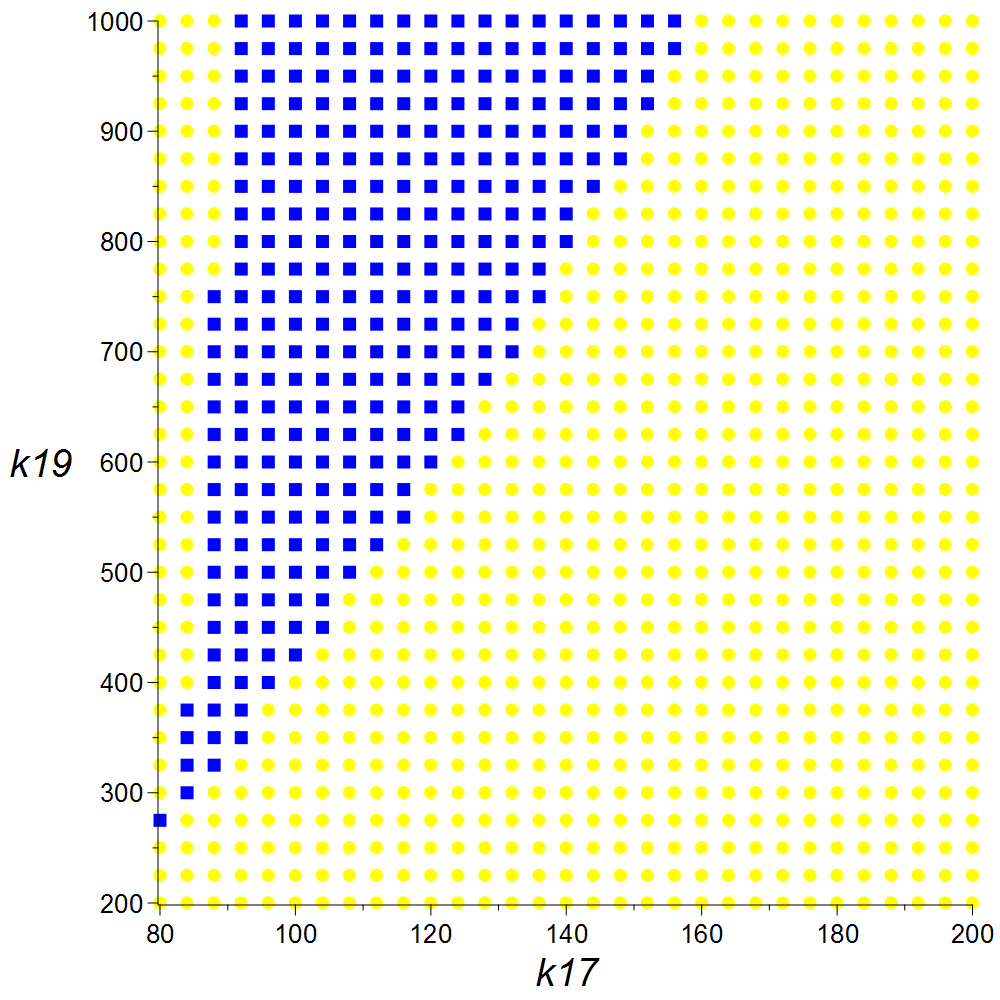}
  \includegraphics[width=0.49\textwidth]{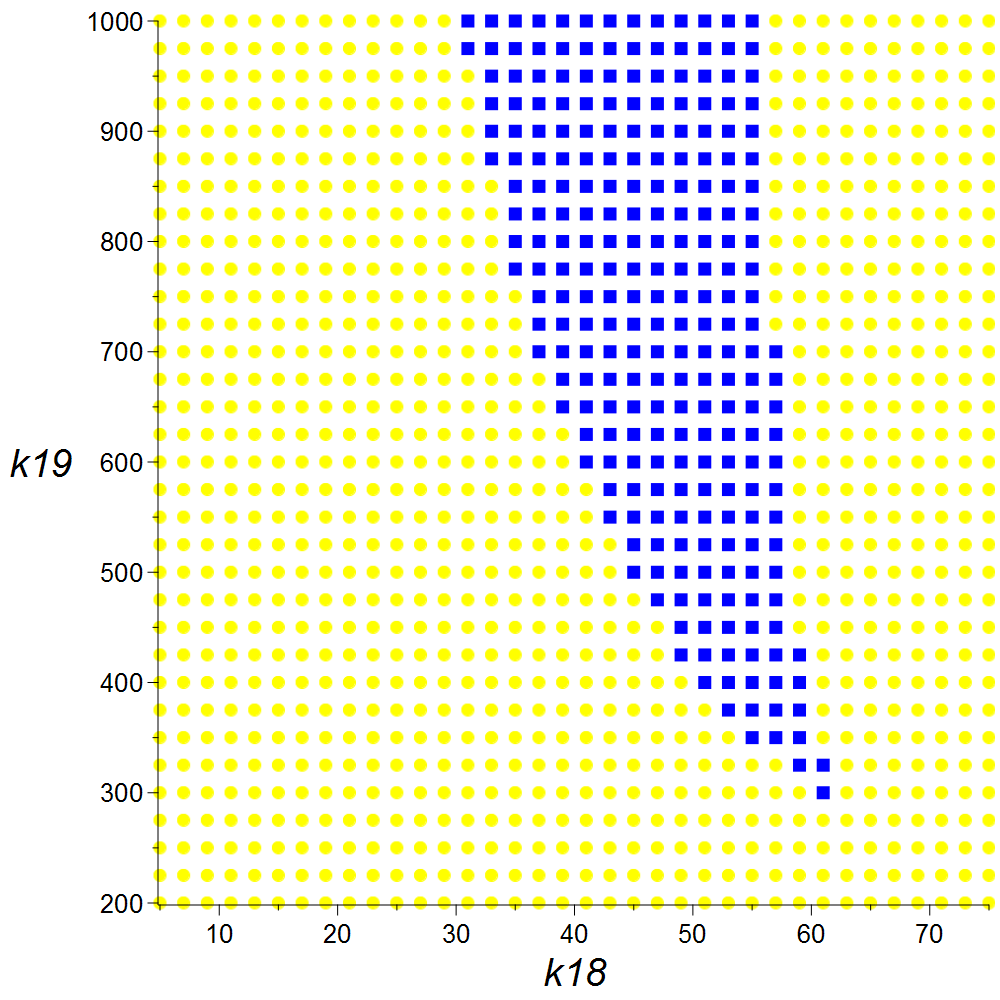}
  \caption{Higher sampling rate for symbolic grid sampling of Model 26.
  \label{FIG:Sys26Detailed}}
\end{figure}

Of course, the grid sampling described in this section scales directly with the number of sample points, so we can easily produce plots with higher sampling rates such as those shown later in Figure \ref{FIG:Sys26Detailed}.

\newpage

\section{Going Further}
\label{SEC:Further}

The work presented is a substantial step forward but there is a wide range of directions for future work.

\subsection{Solution in 3-parameter Space}
\label{SUBSEC:ThreePara}

\begin{figure}[p]
  \centering
  \includegraphics[width=0.49\textwidth]{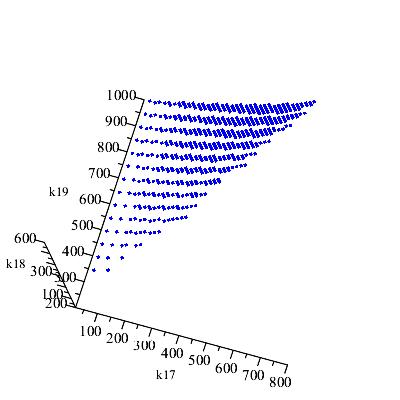}
  \includegraphics[width=0.49\textwidth]{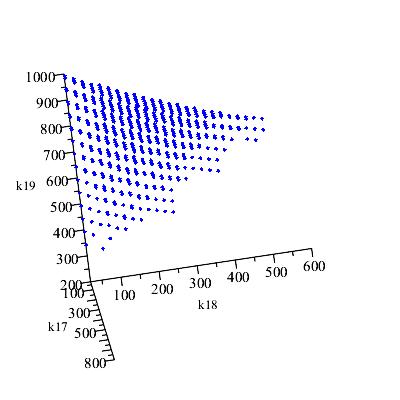}
\caption{3D Maple Point Plot produced grid sampling on Model 26.
\label{FIG:3dPointPlot}}
\end{figure} 

\begin{figure}[p]
  \centering
  \includegraphics[width=0.49\textwidth]{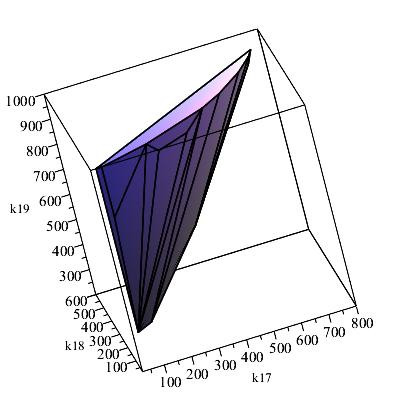}
  \includegraphics[width=0.49\textwidth]{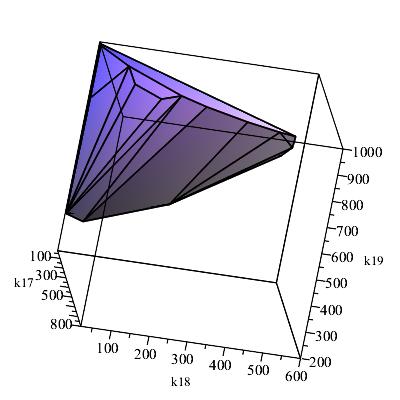}
  \caption{Convex Hull of the bistable points in Figure~\ref{FIG:3dPointPlot} for Model 26\label{FIG:3dConvexHull}}
\end{figure}  

The complexity of the fully symbolic approaches puts a complete analysis over this space out of reach (for now).  However, the grid-sampling method of Section \ref{SEC:vsNumerical} can already be extended into 3 parameters with relative ease: at a cost linearly proportional to the increased number of sample points.  
This was completed for Model 26, where the multistationarity region is bounded on both sides in the $k_{17}$ and $k_{18}$ directions but extends infinitely above in $k_{19}$.  For example, with the $k_{19}$ range bound at 1000 the region is bounded by extending $k_{17}$ to 800 and $k_{18}$ to 600.  With a sample rate of 20 for $k_{17}$ and $k_{18}$ and 50 for $k_{19}$ we have produced a Maple point plot of 20,400 points in 18 minutes. Figure \ref{FIG:3dPointPlot} shows 2D captures of the 3D plot of the bistable points only.
Figure~\ref{FIG:3dConvexHull} gives two views of the convex hull of the bistable points in Figure \ref{FIG:3dPointPlot}.  This was produced using the convex package\footnote{\url{http://www.math.uwo.ca/~mfranz/convex/}}.
We note the lens shape seen in the orientation in the left plot is comparable with the image in the original paper of \citet{Markevich2004} (Fig.~S7).

\subsection{Effect of Other Parameters}
\label{SUBSEC:Pertubation}

Our work has focussed on understanding the behaviour of the system in the 3-parameter space $(k_{17}, k_{18}, k_{19})$ but as described in Section \ref{SEC:Problem} there are many other parameters for which we simply took the values from the BioModels Database.  While there is confidence in the accuracy of these values, an important question for future work is the stability of the approaches we present to small perturbations in these values.

\subsection{Conjecture for Semi-algebraic Solutions without CAD}
\label{SUBSEC:Conjecture}

All our semi-algebraic calculations used CAD as the backend to produce solutions, although after considerable simplification of the input.  CAD is the most expensive technology employed by a significant margin.  Its doubly exponential theoretical complexity is felt clearly in practice and so will be a barrier to studying larger parameter spaces or models.  However, the results of Sections \ref{SEC:Maple} and \ref{SEC:New2Para} hint that the solution could be available without CAD.

Recall from Section \ref{SEC:Maple} that with one free-parameter the key break point in parameter space between 1 and 3 fixed points was determined by a real root of (\ref{EQ:definingpol}), one of the univariate polynomials whose roots were excluded from the validity of the LRT solution component.  Similarly, studying the 28 cells where multistationarity could occur identified in Section \ref{SEC:New2Para} shows that the key region was also identified by the polynomial defining one of the graphs where LRT's solution component was not valid.  

\begin{figure}[p]
\centering
\includegraphics[width=0.45\textwidth]{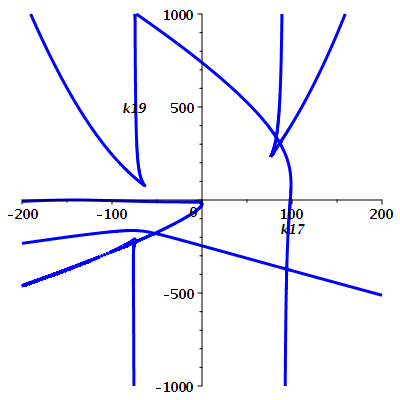}
\includegraphics[width=0.45\textwidth]{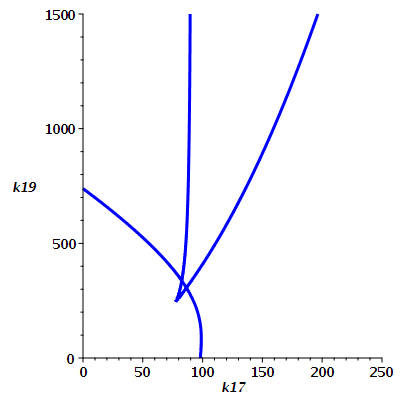}
\caption{Numerical plot of the graph of polynomial (\ref{EQ:Important}) on smaller ranges. \label{FIG:excl1small}}
\end{figure}

\begin{figure}[p]
\centering
\includegraphics[width=0.45\textwidth]{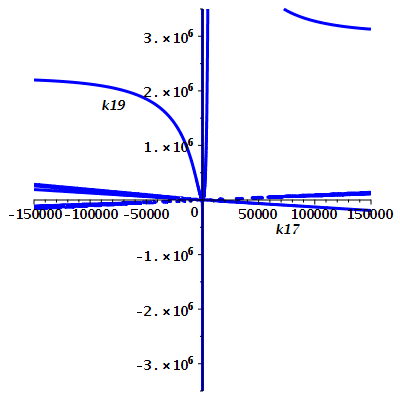}
\includegraphics[width=0.45\textwidth]{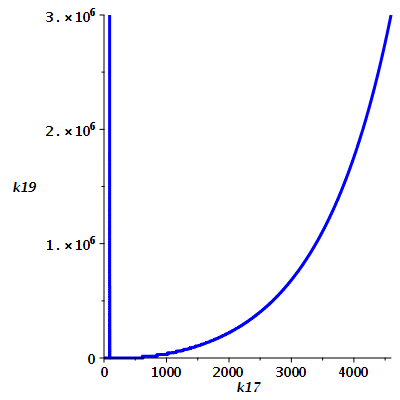}
\caption{Numerical plot of the graph of polynomial (\ref{EQ:Important}) on larger ranges. \label{FIG:excl1big}}
\end{figure}

Figures \ref{FIG:excl1small} and \ref{FIG:excl1big} give numerical plots of the polynomial (\ref{EQ:Important}), the former on smaller ranges and the latter on larger.  The images on the right focus on the upper quadrant of interest and should be compared with Figures \ref{fig:MSRegion1} and \ref{fig:MSRegion2} of the exact multistationarity region.  It is clear that (\ref{EQ:Important}) provides the boundary of this region.  However, as the images on the left show, it is only one segment of the graph of this polynomial that is of interest.  

Of course, this is just an observation.  We have yet to derive a proof that this would always be identified by LRT.  Even if it were there would still be things to clarify:
\begin{itemize}
\item Which polynomial from the several that LRT uses to define excluded regions is the one of interest?  Recall from Section \ref{SEC:Maple} that as well as (\ref{EQ:definingpol}) LRT identified two further polynomials in (\ref{eqtabk19-c3}) and (\ref{eqtabk19-c4}); while in Section LRT identified not only (\ref{EQ:Important}) but also (\ref{EQ:NotImportant}).

\item Which portion of the graph forms the boundary?  The graph of (\ref{EQ:Important}) is a superset of the boundary.  Even, when restricting our view to the positive quadrant (plot on the right of Figure \ref{FIG:excl1small}) there is a second curve segment that does not have relevance to the application.
\end{itemize}
Nevertheless, we have identified a promising conjecture for continued study.  At the least it gives useful insight on where to look for multistationarity without employing CAD.  For example, it could direct future application of detailed grid sampling.

\section{Summary and Final Thoughts}
\label{SEC:Final}

\subsection{Summary}
\label{SUBSEC:Summary}

We have considered the problem of identifying regions of multistationarity in models of biological networks, an important problem with potentially clinical applications. We have investigated a variety of symbolic approaches encompassing multiple algorithms and computer algebra systems.  We have derived semi-algebraic solution formulae and region descriptions for a classic MAPK model; as well as demonstrating the utility of symbolic-numeric grid sampling.  We have drawn together the work first presented at conferences in 2017 \citep{BioISSAC17, BioCASC17} and extended it to give solutions over a 2-parameter space not previously published and a conjecture on where future progress may come from.

\subsection{Final Thoughts}
\label{SUBSEC:FW}

We hope this work will inspire further study on the application of symbolic tools to biological network analysis, from both communities.  Indeed, work on developing Mathematica tools for such problems has now been undertaken by \citet{Lichtblau2017}, inspired by \citet{BioISSAC17} but based on tools for discriminant varieties not considered there.  The study of such real world problems is of great benefit not only to the application domains but also to the software developers: these MAPK studies uncovered bugs in both Regular Chains (see Software Remark \ref{rem:Bug1} in Section \ref{SUBSEC:Maplek19}) and Bertini (see Software Remark \ref{rem:Bug2} in Section \ref{SUBSEC:GSSetup}) which had escaped the numerous other tests and applications of those algorithms.

Key areas of future study include the sensitivity of the analysis to variations in the other parameters (Section \ref{SUBSEC:Pertubation}) and the conjecture described in Section \ref{SUBSEC:Conjecture}.  Additional areas to investigate could include the various degrees of freedom with the algorithms used. For example, we have a free choice of variable ordering: Model 26 has 11 variables corresponding to 39\,916\,800 possible orderings while Model 28 has 16 variables corresponding to
more than $10^{13}$ orderings! Heuristics that exist to help with this choice, such as those of \citet{DSS:04a, BDEW13}, could not discriminate between the orderings on offer, even though the orderings do make a difference to the computation.  Recent work on using machine learning to make such choice by \citet{HEWDPB14, HEDP16} may be applicable. Also, since MAPK problems contain many equational constraints an approach as described by \citet{EBD15} may be applicable for the higher dimensional CADs required to study more parameters.  

Semi-algebraic solutions over 3-parameter space is out of reach at the time of writing.  We note however that instances like MAPK were until recently thought out of reach of symbolic computation altogether, and while writing the ISSAC 2017 contribution we thought the 2-parameter case of Section \ref{SEC:New2Para} out of reach. So further progress will surely follow.

\section*{Acknowledgements}

Section \ref{SEC:Redlog} uses two great free software tools: GNU Parallel for distributing computations on several processors, and yEd for visualization of CAD trees. 

J.~Davenport, M.~England and T.~Sturm are grateful to the European Union's Horizon 2020 Research and Innovation programme, under grant agreement No 712689 (SC\textsuperscript{2}).
H.~Errami, O.~Radulescu, and A.~Weber thanks the French-German Procope-DAAD program for partial support of this research. 
V. Gerdt was partially supported by the RUDN University Program 5-100.  
D.~Grigoriev is grateful to the grant RSF 16-11-10075 and to MCCME for wonderful working conditions and an inspiring atmosphere. 
M.~Ko\v sta has been supported by the DFG/ANR Project STU 483/2-1 SMArT.  

We thank the anonymous reviewers of the present paper and our earlier conference papers for their useful comments which have improved this work.

\vspace*{0.1in}

\section*{Research Data Statement:} Data supporting the research in this paper is freely available in a Zenodo repository: \url{https://doi.org/10.5281/zenodo.2533280}.



\newpage

\appendix

\section{Defining Polynomial of the Section \ref{SEC:Redlog} Break Point}
\label{SEC:AppDefPol}

In Section \ref{REDLOG:k19} a break point where the system moved from 1 to 3 positive real solutions was discovered at around $k_{19} = 409.253$.  The exact point is an algebraic number defined as the only real zero of a polynomial $\sum_{i=0}^{10}c_ik_{19}^i$ with coefficients as below.  
Note that the coefficients are too large to fit on a single line: the line breaks between digits should be read as a continuation of the single coefficient description rather than anything else.  
  \begin{align*}
    c_{10} &= \relax
351590934502740290936895033267017158736060313940693076650\\
&\qquad 155371250411\\
    c_9    &= -2136990728521576742839975277463955832730339831704260805\\
&\qquad 74800781989093156\\
    c_8    &= \relax
253748516412205547742596056350534694325821098839650158040\\
&\qquad 77119110958034090\\
    c_7 &= \relax 
129724930183000227070276392678042592512359916180298528803\\
&\qquad 30004508564391594000\\
    c_6 &= -8468945963692802414226427249726123493448372439778349029\\
&\qquad 355636316929687020660000\\
    c_5 &= \relax
223109827033740645067030166317266433342144083387584862142\\
&\qquad 3683265663846533079600000\\
    c_4 &= -37626500890411225829031917319379205201489948552899492596\\
&\qquad 5885895511831873444245100000\\
    c_3 &= \relax3926210154879086940705799498532015650096895836139617890818\\
&\qquad 0026842806643766783104000000\\
    c_2 &= -249262399074302923497435408127029610630960346245151705777\\
&\qquad 9877596842448287799337600000000\\
    c_1 &= \relax70978850735887473459176997186175978425873267246760023212940\\
&\qquad 616924643171868478080000000000\\
    c_0 &= -106287119283898587694807711492389820499043413890149539483\\
&\qquad 4749613184670362810368000000000000\
  \end{align*}

\section{Polynomial $f(x_1, k_{19})$ from Section \ref{SSSEC:mainSol}}
\label{App:polyF}

In Section \ref{SEC:Maple} we described the application of LRT to (\ref{eq:AllButK19}).  The main solution component provided the formulae \ref{eq:x11Sol}$-$\ref{eq:x2Sol} and required that $f(x_{1}, k_{19}) = \sum_{i=0}^{6} d_ix_{1}^i = 0$ where the coefficients $d_i$ are as given below.

  \begin{align*}
    d_6 &= 16838105723097694257603469\\
    d_5 &= -24078605201553273505077988k_{19} + 7723967969644977896148686580\\
    d_4 &= 8176202638735769127032169k_{19}^2  - 7723411665463544477701499460k_{19} \\
    &\qquad + 1232154357941338876156606812900\\
    d_3 &= 1465408757440589841803452380k_{19}^2  \\
    &\qquad - 798169557586805582842481309800k_{19}  \\
    &\qquad + 83152655240002767729550477640000\\
    d_2 &= 85462524901276846107251669400k_{19}^2  \\
    &\qquad - 35266411401427656834572095140000k_{19}  \\
    &\qquad + 2556805354853318332197489636000000\\
    d_1 &= 1631685649719702672282505500000k_{19}^2  \\
    &\qquad - 721989571100461862477342320000000k_{19}  \\
    &\qquad + 28843755938318780823218400000000000\\
    d_0 &= -7013104139459910876520500000000000k_{19}.
  \end{align*}

\section{Evaluated LRT Solution Component from Section \ref{SEC:New2Para}}
\label{SEC:AppSol}

In Section \ref{SUBSEC:2ParaLRT} we applied LRT to (\ref{eq:M26Red}) to simplify that reduced system further before applying CAD.  The evaluated solution component consisted of the positivity conditions $x_4>0, x_5>0, k_{17}>0, k_{19}>0$ and the two following equations.

\begin{align}
&\big( 
333770827232x_{5}^4 + (3404343829252k_{17} - 6863249873129k_{19} 
\nonumber \\
&\quad - 106111961633240)x_{5}^3 + (  -  3738114656484k_{17}^2 + 7455351062094k_{17}k_{19} 
\nonumber \\
&\quad - 3717236405610k_{19}^2 + 271801037104280k_{17} - 114254579857600k_{19} 
\nonumber \\
&\quad - 831673402560000)x_{5}^2 + ( - 165689075471040k_{17}^2 
\nonumber \\
&\quad + 165225032754600k_{17}k_{19} 
+ 2667668498040000k_{17} - 129311541450000k_{19} 
\nonumber \\
&\quad - 2873589810000000)x_{5} 
- 1835995095480000k_{17}^2 
\nonumber \\
&\quad + 2873589810000000k_{17} \big)x_{4}
\, + \, 
2261223222841x_{5}^5 + ( - 2274797538607k_{17} 
\nonumber \\
&\quad + 2274721722856k_{19} 
 + 174844014037860)x_{5}^4 + (13574315766k_{17}^2 
 \nonumber \\
&\quad - 27072815781k_{17}k_{19} 
 + 13498500015k_{19}^2 - 176205245392020k_{17} 
 \nonumber \\
&\quad - 883400777350k_{19} 
 + 6648403506290000)x_{5}^3 + (1361231354160k_{17}^2 
 \nonumber \\
&\quad - 1355303940900k_{17}k_{19} 
 - 6671855445710000k_{17} + 6724440511425000k_{19} 
 \nonumber \\
&\quad + 149432011365000000)x_{5}^2 + (23451939420000k_{17}^2 
\nonumber \\
&\quad  - 149432011365000000k_{17})x_{5} = 0 
\label{eq:withX4}
\end{align}

\begin{align}
&487656080889027413x_{5}^6 
 +  ( - 1352408212353388839k_{17} 
\nonumber \\
&\quad + 2227511326365959821k_{19} + 97141513552593345960)x_5^5 
\nonumber \\
&\quad 
+ (1810515745366146214k_{17}^2 - 4490852292185431392k_{17}k_{19} 
\nonumber \\
&\quad + 2680336546819285178k_{19}^2 - 220676803454346691680k_{17} 
\nonumber \\
&\quad + 166893970054477098860k_{19} + 6819142839866322930800)x_5^4
\nonumber \\
&\quad +  ( - 945763613901784788k_{17}^3 + 2832008529145922346k_{17}^2k_{19} 
\nonumber \\
&\quad - 2826726216586490328k_{17}k_{19}^2 + 940481301342352770k_{19}^3 
\nonumber \\
&\quad + 239398211250170709480k_{17}^2 - 397099010517367066520k_{17}k_{19} 
\nonumber \\
&\quad + 89401058522195274400k_{19}^2 - 14716205773190097360400k_{17} 
\nonumber \\
&\quad + 8313128696476184347000k_{19} + 308330512782039741800000)x_5^3
\nonumber \\
&\quad  +  ( - 115862921348417363760k_{17}^3 + 231195450091661030160k_{17}^2k_{19} 
\nonumber \\
&\quad - 115332528743243666400k_{17}k_{19}^2 + 11639096756278536898400k_{17}^2 
\nonumber \\
&\quad - 8542395106508656744000k_{17}k_{19} + 523361626689201300000k_{19}^2 
\nonumber \\
&\quad - 420660564631403190200000k_{17} + 15948686720945888000000k_{19} 
\nonumber \\
&\quad + 5159677297706895600000000)x_5^2
 +  ( - 3742033822954762468800k_{17}^3 
\nonumber \\
&\quad + 3732854354558173572000k_{17}^2k_{19} + 148648818114128214000000k_{17}^2 
\nonumber \\
&\quad - 26235555941563878000000k_{17}k_{19} - 5484239465944512000000000k_{17} 
\nonumber \\
&\quad + 5101447069138124250000000k_{19} + 113365490425291650000000000)x_5
 \nonumber \\
&\quad -  36318766264764765600000k_{17}^3 + 324562168237616400000000k_{17}^2 
\nonumber \\
&\quad - 113365490425291650000000000k_{17} = 0
\label{eq:withoutX4}
\end{align}

\newpage

\section{The polynomials in $(k_{17}, k_{19}$)-space excluded by LRT in Section \ref{SEC:New2Para}}
\label{SEC:AppExcl}

The evaluated solution component in the previous appendix is guaranteed to describes the solution everywhere except upon the graphs of two polynomials in $(k_{17}, k_{19}$)-space.  
The smaller of these polynomials is as follows:
\begin{align}
&306149569674418411007002633445069482118718951168k_{17}^5
\label{EQ:NotImportant} \\ &\quad -928141594350529690019570716839242728610620920576k_{17}^4k_{19}
\nonumber \\ &\quad+949816997057955538346464679473943447453989559073k_{17}^3k_{19}^2
\nonumber \\ &\quad -339807489761995650662227742210436637550992161090k_{17}^2k_{19}^3
\nonumber \\ &\quad+11982517380151391328331146130666436588904571425k_{17}k_{19}^4
\nonumber \\ &\quad
-48999080739606236406966583535007903157444819975616k_{17}^4
\nonumber \\ &\quad+132280370740212793297769000628045387812057010666000k_{17}^3k_{19}
\nonumber \\ &\quad-206266836118899613221788680523164250210223905969850k_{17}^2k_{19}^2
\nonumber \\ &\quad+107105747411519378668353959818922318524218807524875k_{17}k_{19}^3
\nonumber \\ &\quad-25449048291062715282099864289265288529894455756250k_{19}^4
\nonumber \\ &\quad+2851566891087903587412213909599967256213769704859200k_{17}^3
\nonumber \\ &\quad-9096628139611598903423536369544450313430913324700000k_{17}^2k_{19}
\nonumber \\ &\quad+8738534807301297185258048178125213648416011272272500k_{17}k_{19}^2
\nonumber \\ &\quad-4707089603080633815275363638970588496447978811156250k_{19}^3
\nonumber \\ &\quad-154536715731414742272245150527717608235719602790480000k_{17}^2
\nonumber \\ &\quad+337583233182458249596138053094849235485707240504000000k_{17}k_{19}
\nonumber \\ &\quad-419058873458723903282123960357587776939186070160625000k_{19}^2
\nonumber \\ &\quad+4055778459605626549669861788992643508030535903264000000k_{17}
\nonumber \\ &\quad-10550282279371566387655279963112364142636872990000000000k_{19}
\nonumber \\ &\quad-80103658453495029562086963732044424664873830868000000000. \nonumber
\end{align}
The larger is defined by 
\begin{equation}
\sum_{i=0}^{14} e_ik_{17}
\label{EQ:Important}
\end{equation}
where the $e_{i}$ are univariate polynomials in $k_{19}$ given over the following pages.  In Section \ref{SUBSEC:Conjecture} we noted that part of the graph of this polynomial forms the boundary of the desired region in $(k_{17}, k_{19}$)-space where multiple solutions exist.

\includepdf[pages=-, landscape=true]{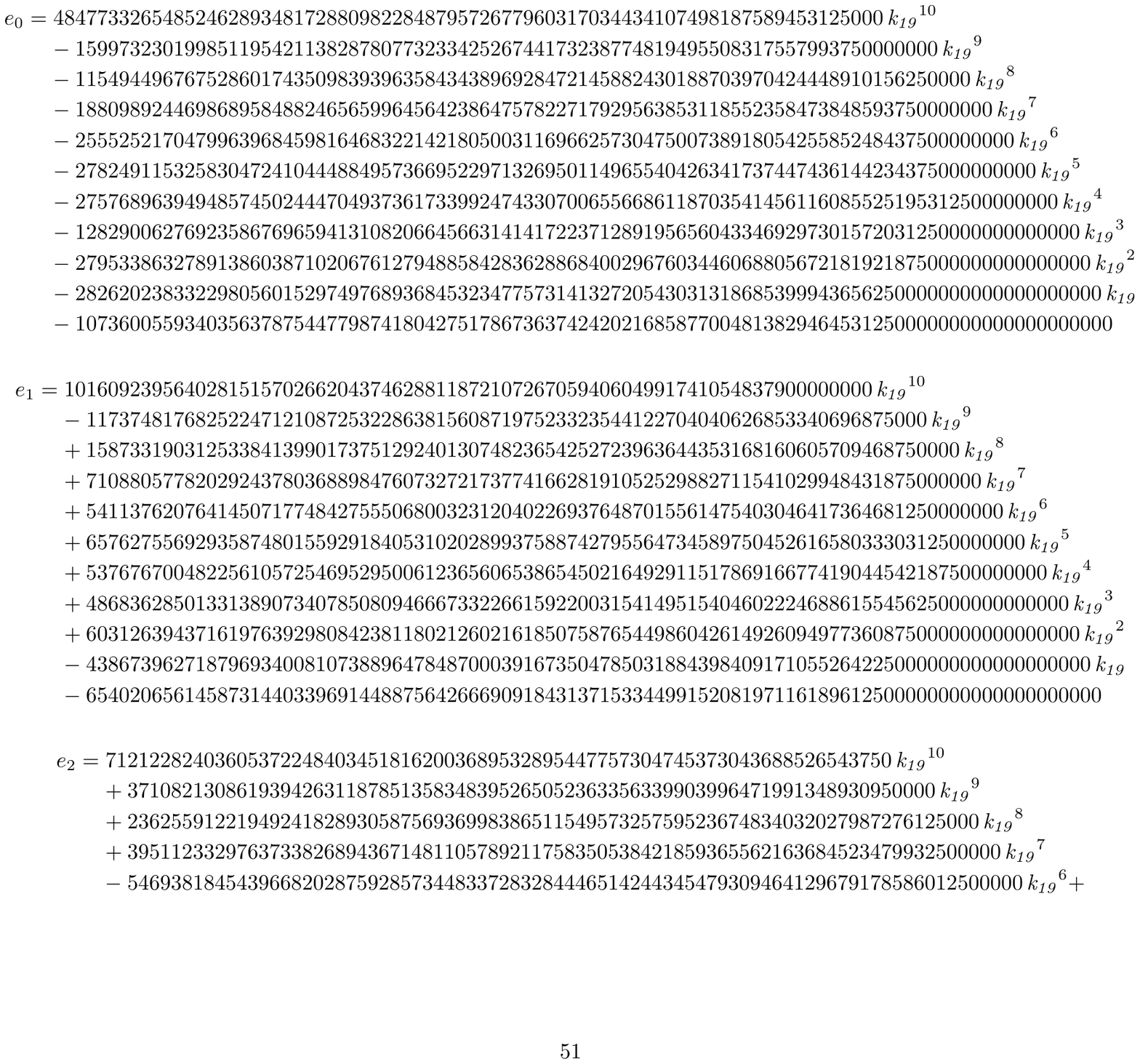}




\vspace*{0.1in}


\bibliographystyle{elsarticle-harv} 
\biboptions{authoryear}
\bibliography{BioBib}

\end{document}